\documentclass[review, 3p, 12pt]{elsarticle}
\usepackage{soul}
\usepackage{graphicx, graphics}
\usepackage{syntonly}
\usepackage{amssymb, amsmath, amsfonts}
\usepackage{subfigure}
\usepackage{multirow}
\usepackage{color}
\usepackage{bm}

\newcommand{\bx}{\bm X}

\providecommand{\br}{\bm r}
\providecommand{\bv}{\bm v}
\providecommand{\bK}{\bm K}
\providecommand{\cS}{\mathcal S}

\providecommand{\tnabla}{\tilde{\nabla}}

\newcommand{\pfrac}[2]{\frac{\partial #1}{\partial #2}}

\newcommand{\tr}[1]{\mbox{tr}(#1)}

\begin{document}
\begin{frontmatter}
\title{Ductile fracture modeling by phase field, Hencky strain elasticity and finite J2 plasticity using nonlocal operator method}
\author[HU]{Huilong Ren}
\ead{huilong.ren@iop.uni-hannover.de}
\author[BU]{Timon Rabczuk}
\ead{timon.rabczuk@uni-weimar.de}
\author[HU,TU]{Xiaoying Zhuang\corref{cor1}}
\ead{zhuang@iop.uni-hannover.de}

\address[HU]{Institute of Photonics, Department of Mathematics and Physics, Leibniz University Hannover,Germany}
\address[BU]{Institute of Structural Mechanics, Bauhaus-Universit at Weimar, Germany}
\address[TU]{State Key Laboratory of Disaster Reduction in Civil Engineering, College of Civil Engineering,Tongji University, Shanghai 200092, China}
\cortext[cor1]{Institute of Photonics, Department of Mathematics and Physics, Leibniz University Hannover,Germany. zhuang@iop.uni-hannover.de;}

\begin{abstract}
A phase field model for ductile fracture considering Hencky strain and finite J2 plasticity is presented using the nonlocal operator method. A variational derivation of J2 plasticity at finite strain with a phase field model is performed. The method includes a logarithmic strain tensor and an exponential mapping in the plasticity evolution. A spectral decomposition based algorithm for computing the first and second order derivatives of the composite matrix function is implemented. A consistent tangential stiffness matrix is derived and used in Newton-Raphson iterations. Several numerical examples are performed to validate the method, including notched single-edged plates with brittle fracture or ductile fracture and necking of a bar with/without phase field model.
\end{abstract}
\end{frontmatter}

%

\section{Introduction}\label{sec:introduction}
\vspace{-2pt}

Safety evaluation of metallic materials or structures is important for many engineering applications. In the process of determining the critical state of the structures, the formation of plastic deformation and the development of brittle/ductile fractures are often involved. Plastic deformation is the ability of a solid material to permanently deform under the action of external forces when the critical state of elastic deformation reaches yielding. Ductile fracture with finite deformation means that the fracture develops in conjunction with the permanent deformation and the possible thermal effect. The simulation of ductile fracture encounters two difficulties: the finite elastoplastic deformation and the evolution of the fractures. 

The mechanism of elastoplastic deformation is closely related to micromechanics based on the internal structure of materials. However, a phenomenological theory of plasticity based on a macroscopic description is useful for many engineering applications. Finite strain plasticity is usually based on the multiplicative decomposition of the deformation gradient proposed by Lee \cite{lee1968finite,lee1969elastic}. The multiplicative decomposition consists of an elastic and a plastic part of the deformation tensor and allows a material frame invariant description of plasticity \cite{hill1979aspects}. The thermodynamically consistent finite strain plasticity framework has been well presented by Havner, Hill, Rice, and others \cite{fardshisheh1974representation,rice1975continuum,asaro1983crystal,nemat1982finite,dafalias1987issues}. 

The plastic model mainly considers plastic deformation, and the evolution of sharp discontinuities such as fractures requires additional techniques such as the phase field model. Phase field methods have proven to be very successful in modeling brittle fractures and ductile fractures. Based on the Griffith energy principle, the fracture surface is represented by a phase field energy functional in terms of $\Gamma$-convergence. The evolution of the phase field reflects the automatic determination of the fracture direction and evolution. The development of the phase field energy framework took advantage of the Mumford-Shah functional \cite{mumford1985boundary} and the Ambrosio-Tortorelli functional \cite{ambrosio1990approximation}. Mumford-Shah functional was introduced in \cite{mumford1985boundary} for image segmentation, which is written as
\begin{align}
\min _{u, K}\left\{\alpha \int_{\Omega}|u-\bar{u}|^2+\int_{\Omega \backslash K}|\nabla u|^2+\beta \mathcal{H}^{n-1}(K)\right\},
\end{align}
where $u \in\Omega \backslash K$, $\bar{u}$ the prescribed field value of $u$, $K \subset \Omega$ closed in $\Omega, \mathcal{H}^{n-1}(K)<\infty$, where $\mathcal{H}^{n-1}$ is the ($n-1$)-dimensional Hausdorff measure (e.g. surface area in 3D when $n=3$), and $\alpha, \beta>0$ are fixed parameters. The functional consists of the sharp interface in domain $K$ and low gradient terms in $\Omega \backslash K$. In Ref \cite{ambrosio1990approximation}, Ambrosio and Tortorelli proposed an elliptic functional (i.e.  the Ambrosio-Tortorelli functional) as follows:
\begin{align}
A T_\epsilon(u, c)=& \alpha \int_{\Omega}|u-\bar{u}|^2 d V+\int_{\Omega} c^2|\nabla u|^2 d V+\beta {\int_{\Omega}\left(\epsilon|\nabla c|^2+\frac{(c-1)^2}{4 \epsilon}\right) d V},\label{eq:ATe}
\end{align}
where $\epsilon$ is a positive fixed parameter. The authors proved that $AT_{\epsilon}(u,c)$ converges in the sense of $\Gamma$-convergence with the Mumford-Shah functional when $\epsilon\to 0$. In the spirit of the Ambrosio-Tortorelli functional, Francfort and Marigo \cite{francfort1998revisiting} replaced the gradient term $|\nabla u|^2$ in Eq.\ref{eq:ATe} by the elastic strain energy density in solid mechanics and formulated the brittle fracture as an energy minimization problem. The numerical experiments were performed in \cite{bourdin2000numerical}. The additional field $c$ greatly simplifies the description of the fracture. In Ref \cite{miehe2010thermodynamically}, Miehe etal developed the thermodynamically consistent phase field model based on the spectral decomposition of the strain tensor in isotropic elastic material, showing great stability and capability in complicated crack simulations. Phase field models have been used to solve brittle fractures \cite{karma2001phase,ambati2015review}, hydraulic fractures \cite{wilson2016phase,miehe2015minimization,zhou2019phase,heider2021review}, ductile fractures with plasticity \cite{borden2016phase,badnava2017phase,dittmann2018variational,haveroth2020non,aldakheel2019virtual}, geomaterials \cite{choo2018coupled,fei2021double,bryant2018mixed,you2020incorporation}, plate and shell \cite{areias2016phase,proserpio2021phase}, multiphysics problems \cite{miehe2015phase,miehe2015phase2,dittmann2020phase} and many others. For a more comprehensive review of the phase field model, the reader is referred to \cite{wu2020phase,bui2021review,zhuang2022phase}.




Numerical methods for the phase field model are usually based on the finite element method \cite{miehe2010thermodynamically} or isogeometric analysis \cite{borden2014higher,chen2020adaptive}. Nonlocal Operator Method (NOM) is a numerical method based on nonlocality to solve partial differential equations \cite{rabczuk2019nonlocal,ren2020nonlocal,ren2020nonlocalgrad,ren2020higher}. NOM as a generalization of the dual-horizon Peridynamics (PD) \cite{Ren2015}, extends the first-order nonlocal gradient to the higher-order nonlocal derivatives without recourse to shape functions. By adopting a variational framework and weighted residual methods, NOM is applicable to many physical problems based on the weak form \cite{Ren2021May}. The implementation of NOM allows an inhomogeneous discretization and has no restriction on the support size and support shape. Meanwhile, it has the capability of deriving the nonlocal models based on the traditional local forms or the weak forms \cite{Ren2021Sep}. It has been applied to solve various physical problems such as linear/nonlinear gradient elasticity \cite{ren2020nonlocalgrad}, waveguide problem in electromagnetic field \cite{rabczuk2019nonlocal}, Cahn-Hilliard equation \cite{Ren2021May}, von-Karmman equations for thin plates \cite{ren2020higher}. In this work, we apply the NOM to model ductile fracture due to plasticity with consideration of both material and geometric nonlinearity. 

The remainder of the paper is organized as follows. In Section 2, the first- order NOM is briefly discussed and the matrix form, the tangent stiffness matrix of the hourglass energy, are explicitly formulated with a discrete setting of the support domain. In Section 3, the spectral decomposition of the matrix function is presented and its first-order and second- order derivatives are discussed. In Section 4, the governing equation of Hencky strain elasticity is derived and the residual vector and tangent stiffness matrix are derived. In Section 5, the theoretical framework of multiplicative J2 elastoplasticity including flowing rule and yield function is presented. In Section 6, the phase field J2 plasticity model is discussed. Section 7 presents several numerical examples, including a notched-single-edge plate with brittle fracture or ductile fracture and the necking of a bar with/without ductile fracture, to validate the NOM scheme. Some conclusions are drawn in Section 8.

\section{Brief review of nonlocal operator method}\label{sec:nommethod}

\subsection{Support, dual-support and nonlocal operators}\label{sec:support}
\begin{figure}[htp]
\centering
\subfigure[]{
\label{fig:Coord}
\includegraphics[width=.4\textwidth]{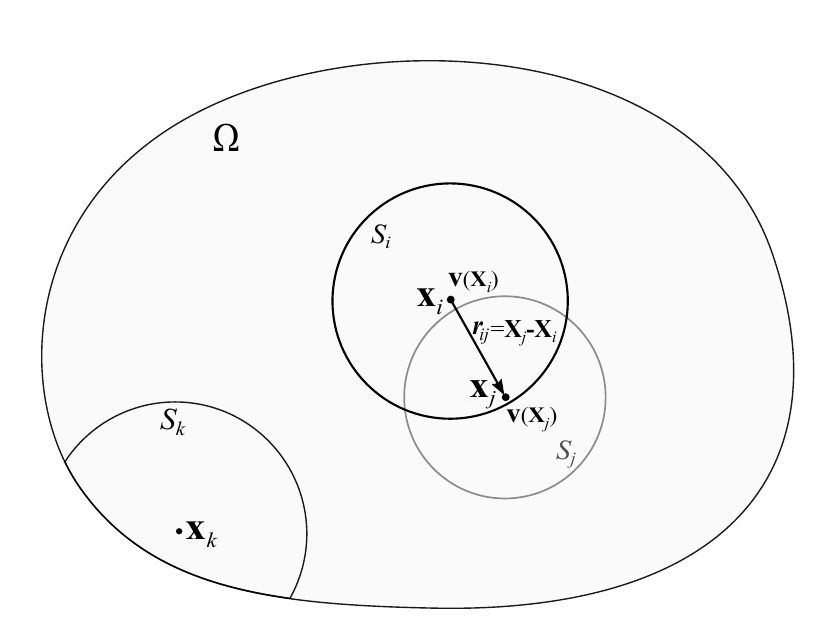}}
\subfigure[]{
\label{fig:4support}
\includegraphics[width=.35\textwidth]{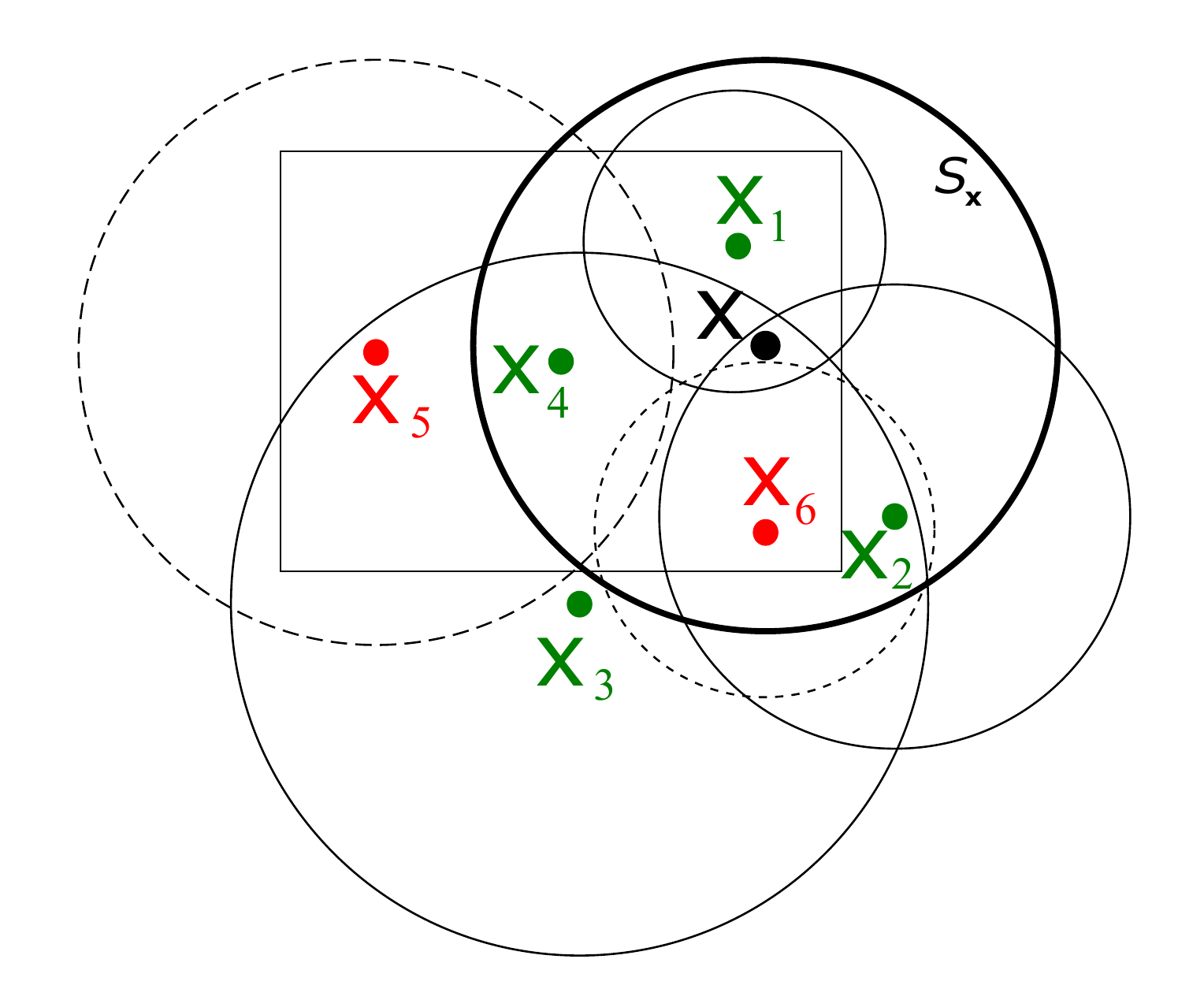}}\\
\caption{(a) Domain and notations. (b) Schematic diagram for support and dual-support, all the above formes are support areas, $\{\bx_1,\bx_2,\bx_4,\bx_6\}\subset \cS_{\bx} $, $\{\bx_1,\bx_2,\bx_3,\bx_4\} \subset \cS_{\bx}'$.}
\end{figure}
Consider a domain as shown in Fig.\ref{fig:Coord}, let $\bx$ be spatial coordinates in the initial configuration $\mathbf \Omega$; $\br_{ij}:=\bx_{j}-\bx_{i}$ is a vector (or a spatial vector, or simply a vector) starts from $\bx_i$ to $\bx_j$; $\bv_i:=\bv(\bx_i,t)$ and $\bv_j:=\bv(\bx_j,t)$ are the field values for $\bx_i$ and $\bx_j$, respectively; $\bv_{ij}:=\bv_j-\bv_i$ is the relative field vector for spatial vector $\br_{ij}$.

\textbf{Support} $\cS_{i}$ is a finite-size neighborhood of point $\bx_i$. {Support} $\cS_{i}$ {specifies the range of nonlocal interaction that happened with respect to point} $\bx_i$. The main function of a support is to define different nonlocal operators. In mathematics, a specific quantitative measure of the support can be described by a moment (or a shape tensor),
\begin{align}
\bK_{i}:=\int_{\cS_{i}}w(\br_{ij}) \br_{ij} \otimes \br_{ij} d V_{j},
\end{align}
where $w(\br_{ij})$ is the weight function.

\textbf{Dual-support} is defined as a union of the points whose supports include $\bx_i$, denoted by
\begin{align}
\cS_{i}'=\{\bx_j|\bx_i \in \cS_{j}\} \label{eq:dualsupport}.
\end{align}
The point $\bx_j$ forms dual-vector $\br_{ij}'(=\bx_i-\bx_j=-\br_{ij})$ in $\cS_{i}'$. On the other hand, $\br_{ij}'$ is the spatial vector formed in $\cS_{j}$. One example to illustrate the range of support and dual-support is shown in Fig.\ref{fig:4support}. Seven representative points and their support areas are drawn in the figure. After defining the support area, we can see that the support of $\bx$ contains $\{\bx_1,\bx_2,\bx_4,\bx_6\}$, while the dual-support of $\bx$ consists of $\{\bx_1,\bx_2,\bx_3,\bx_4\}$.

\subsection{Nonlocal operators in support}\label{ssec:nos}
The common operators in calculus include the gradient of the scalar and vector field, the curl and divergence of the vector field. {The definitions of some nonlocal operators can be found in reference} \cite{du2013nonlocal}. These operators have the corresponding nonlocal forms based on the Taylor series expansion. We use ${\tnabla}$ {to denote the nonlocal operator, while the local operators follow the conventional notations.

The first-order nonlocal operator for vector-valued field $\bm u$ at point $\bm x_i$ is defined as
\begin{align}
\tilde{\nabla} * \bm u_i:=\int_{\mathcal S_i} \omega(r_{ij}) (\bm K_i^{-1} \bm r_{ij}) * \bm u_{ij} d V_j,\label{eq:tnablaA}
\end{align}
where $*\in \{\cdot, \otimes, \times\}$ denote the inner product, tensor product and cross product, respectively. The continuous form is inconvenient for implementation. After discretization of the domain by particles, the whole domain is represented by
\begin{align}
\Omega=\sum_{i=1}^{N} V_i,
\end{align}
where $i$ is the global index of volume $ V_i$, $N$ is the number of particles in $\Omega$.

Particles in $\cS_{i}$ are represented by
\begin{align}
\mathcal N_{i}=\{i,j_1,..,j_k,..,j_{n_{i}}\},\label{eq:Ni}
\end{align}
where $j_1,..,j_k,..,j_{n}$ are the global indices of neighbors of particle $i$. Herein, the number of particles in each support is fixed as $n$, although $n$ can vary for each particle.

The nonlocal gradient for scalar-valued field $u$ at point $\bm X_i$ in continuous form and discrete form is
\begin{align}
\tnabla u_i=\int_{\mathcal S_i} \omega(r_{ij}) \bm K_i^{-1} \bm r_{ij}\, (u_j-u_i) d V_j=\sum_{j\in \mathcal N_i} V_j \omega(r_{ij}) \bm K_i^{-1} \bm r_{ij}\, (u_j-u_i). \label{eq:nablau}
\end{align}
Let
\begin{align}
V_j \omega(r_{ij}) \bm K_i^{-1} \bm r_{ij}=\begin{cases} (r^x_j, r^y_j) \mbox{ in 2D}\\
(r^x_j, r^y_j,r^z_j) \mbox{ in 3D.}\\
\end{cases}
\end{align}
Then $\tnabla u_i$ can be written as matrix form as
\begin{align}
\tnabla u_i=\begin{pmatrix}
\pfrac{u_i}{x}\\
\pfrac{u_i}{y}\\
\pfrac{u_i}{z}
\end{pmatrix}=\underbrace{\begin{pmatrix}
-\sum_{j\in \mathcal N_i} r^x_j & r^x_{j_1} & \dots & r^x_{j_n} \\
-\sum_{j\in \mathcal N_i} r^y_j & r^y_{j_1} & \dots & r^y_{j_n} \\
-\sum_{j\in \mathcal N_i} r^z_j & r^z_{j_1} & \dots & r^z_{j_n} \\
\end{pmatrix}}_{\bm B_i}\cdot \underbrace{\begin{pmatrix}
u_i\\
u_{j_1}\\
\vdots\\
u_{j_n}
\end{pmatrix}}_{\bm U_i}.
\end{align}
The matrix form of $\tnabla u_i$ is similar to the gradient of the shape function in the finite element method.

The energy functional for scalar-valued field $u$ at point $\bx_i$
\begin{align}
\mathcal F^{hg}&=\frac 12\frac{\alpha}{\mbox{tr}(\bm K_i)}\int_{\mathcal S_i} \omega(r_{ij}) (\tnabla u_i \bm r_{ij}-u_{ij})\cdot (\tnabla u_i \bm r_{ij}-u_{ij}) d V_j\notag\\
&=\frac 12\frac{\alpha}{\mbox{tr}(\bm K_i)}\sum_{j\in\mathcal N_i} \omega(r_{ij}) (\tnabla u_i \bm r_{ij}-u_{ij})\cdot (\tnabla u_i \bm r_{ij}-u_{ij}) V_j,
\end{align}
where $\alpha$ is a penalty coefficient. $\mbox{tr}(\square)$ denotes the trace of the matrix.

Replacing $\nabla u_i$ with $\tnabla u_i$ in Eq.\ref{eq:nablau}, $\mathcal F^{hg}$ in discrete form can be simplified as
\begin{align}
\mathcal F^{hg}&= \frac 12 \bm U_i^T\mathbb K_i^{hg} \bm U_i,
\end{align}
where
\begin{align}
\mathbb K_i^{hg}:=\frac{\alpha}{\mbox{tr}(\bm K_i)} \Big(
\begin{pmatrix}
\sum_{j\in \mathcal N_i} \omega(r_{ij}) V_j & -\omega(r_{ij_1}) V_{j_1} & \cdots & -\omega(r_{ij_n}) V_{j_n}\\
-\omega(r_{ij_1}) V_{j_1} & \omega(r_{ij_1}) V_{j_1} &\bm 0 & 0\\
\vdots & \bm 0 & \ddots & \bm 0\\
-\omega(r_{ij_n}) V_{j_n} & 0 & \bm 0 & \omega(r_{ij_n}) V_{j_n}
\end{pmatrix}-\bm B_i^T \bm K_i \bm B_i\Big).\label{eq:khg1}
\end{align}
Actually, $\mathbb K_i^{hg}$ is the hourglass matrix that can suppress the hourglass mode when adding to the tangent stiffness matrix of the physical model. More details can be found in \cite{ren2020nonlocal}.
 
Eq.\ref{eq:khg1} is for a single field $u$. For vector-valued field $\bm u=(u,v,w)$ in 3-dimensional space, the nonlocal derivatives in matrix form are
\begin{align}
\begin{pmatrix}
\pfrac{u_i}{x}\\
\pfrac{u_i}{y}\\
\pfrac{u_i}{z}\\
\pfrac{v_i}{x}\\
\pfrac{v_i}{y}\\
\pfrac{v_i}{z}\\
\pfrac{w_i}{x}\\
\pfrac{w_i}{y}\\
\pfrac{w_i}{z}\\
\end{pmatrix}=\underbrace{\begin{pmatrix}
\text{-}\sum_{j\in \mathcal N_i} r^x_j ,\, 0 ,\, 0 ,\, r^x_{j_1} ,\, 0 ,\, 0,\, \cdots ,\, r^x_{j_n} ,\, 0 ,\, 0 \\
\text{-}\sum_{j\in \mathcal N_i} r^y_j ,\, 0 ,\, 0,\, r^y_{j_1} ,\, 0 ,\, 0,\, \cdots ,\, r^y_{j_n} ,\, 0 ,\, 0 \\
\text{-}\sum_{j\in \mathcal N_i} r^z_j ,\, 0 ,\, 0,\, r^z_{j_1} ,\, 0 ,\, 0 ,\, \cdots ,\, r^z_{j_n} ,\, 0 ,\, 0 \\
0,\, \text{-}\sum_{j\in \mathcal N_i} r^x_j ,\, 0 ,\, 0 ,\, r^x_{j_1} ,\, 0 ,\, \cdots,\, 0 ,\, r^x_{j_n} ,\, 0 \\
0,\, \text{-}\sum_{j\in \mathcal N_i} r^y_j ,\, 0 ,\, 0,\, r^y_{j_1} ,\, 0 ,\, \cdots ,\, 0 ,\, r^y_{j_n} ,\, 0 \\
0,\, \text{-}\sum_{j\in \mathcal N_i} r^z_j ,\, 0 ,\, 0,\, r^z_{j_1} ,\, 0 ,\, \cdots,\, 0 ,\, r^z_{j_n} ,\, 0 \\
0 ,\, 0,\, \text{-}\sum_{j\in \mathcal N_i} r^x_j ,\, 0 ,\, 0 ,\, r^x_{j_1} ,\, \cdots ,\, 0 ,\, 0,\, r^x_{j_n} \\
0 ,\, 0,\, \text{-}\sum_{j\in \mathcal N_i} r^y_j ,\, 0 ,\, 0,\, r^y_{j_1} ,\, \cdots ,\, 0 ,\, 0 ,\, r^y_{j_n} \\
0 ,\, 0,\, \text{-}\sum_{j\in \mathcal N_i} r^z_j ,\, 0 ,\, 0,\, r^z_{j_1} ,\, \cdots ,\, 0 ,\, 0,\, r^z_{j_n} \\
\end{pmatrix}}_{\bm B_i}\underbrace{\begin{pmatrix}
u_i\\v_i\\w_i\\u_{j_1}\\v_{j_1}\\w_{j_1}\\\vdots\\v_{j_n}\\w_{j_n}
\end{pmatrix}}_{\bm U_i}.\label{eq:BU3}
\end{align}

The corresponding energy functional at point $\bx_i$
\begin{align}
\mathcal F^{hg}&=\frac 12\frac{\alpha}{\mbox{tr}(\bm K_i)}\int_{\mathcal S_i} \omega(r_{ij}) (\tnabla \bm u_i \bm r_{ij}-\bm u_{ij})\cdot (\tnabla \bm u_i \bm r_{ij}-\bm u_{ij}) d V_j\notag\\
&=\frac 12\frac{\alpha}{\mbox{tr}(\bm K_i)}\sum_{j\in\mathcal N_i} \omega(r_{ij}) (\tnabla \bm u_i \bm r_{ij}- \bm u_{ij})\cdot (\tnabla \bm u_i \bm r_{ij}- \bm u_{ij}) V_j,
\end{align}

Replacing $\nabla \bm u_i$ with $\tnabla \bm u_i$ in Eq.\ref{eq:tnablaA}, $\mathcal F^{hg}$ in discrete form can be simplified as
\begin{align}
\mathcal F^{hg}&= \frac 12 \bm U_i^T\mathbb K_i^{hg} \bm U_i,
\end{align}
where $\bm U_i^T=(u_i,v_i,w_i,u_{j_1},v_{j_1},w_{j_1},\cdots,u_{j_n},v_{j_n},w_{j_n})$, and $\mathbb K_i^{hg}$ is
the hourglass tangent stiffness matrix for vector field $\bm u=(u,v,w)$:
\begin{align}
\mathbb K_i^{hg}:=\frac{\alpha}{\mbox{tr}(\bm K_i)} \Big(
\begin{pmatrix}
\sum_{j\in \mathcal N_i} \omega(r_{ij}) V_j & \text{-}\omega(r_{ij_1}) V_{j_1} & \cdots & \text{-}\omega(r_{ij_n}) V_{j_n}\\
\text{-}\omega(r_{ij_1}) V_{j_1} & \omega(r_{ij_1}) V_{j_1} &\bm 0 & 0\\
\vdots & \bm 0 & \ddots & \bm 0\\
\text{-}\omega(r_{ij_n}) V_{j_n} & 0 & \bm 0 & \omega(r_{ij_n}) V_{j_n}
\end{pmatrix}\otimes \bm I-\bm B_i^T (\bm K_i\otimes \bm I) \bm B_i\Big).\label{eq:khg3}
\end{align}
Herein, only the first order nonlocal gradient is provided. For higher order formulation of NOM, the interested reader is referred to \cite{ren2020higher}.

\section{Matrix function and spectral decomposition}\label{sec:matrixFun}

A difficulty in single crystal plasticity with large strains is the calculation of exponential matrix functions and their derivatives for the incompressibility of the plasticity evolution. The use of exponential matrix function allows the incompressibility of plasticity evolution if the exponential matrix function is calculated accurately. There are several methods of calculating matrix functions, for example, direct summation of (truncated) series expansion of exponential matrix \cite{ortiz2001computation}, spectral decomposition \cite{carlson1986derivative}, generating function of tensor functions \cite{lu2004exact,korelc2014closed} and so on. Lu \cite{lu2004exact} used stem function to compute the arbitrary tensor functions and their first derivatives with closed-form, singularity-free expressions. The spectral decomposition \cite{ortiz2001computation,carlson1986derivative} can provide a closed-form representation of both the exponential and its derivative. In the case of repeated eigenvalues, special care should be taken to avoid numerical singularities. The spectral decomposition has been used in several implementations of plasticity at finite-strain plasticity \cite{ibrahimbegovic1999finite,ortiz2001computation}. Korelc and Stupkiewicz \cite{korelc2014closed} have proposed a closed-form matrix function using automatic differentiation of an appropriate scalar generating function. High numerical accuracy is achieved even for the repeated eigenvalues. In the present work, we use spectral decomposition to compute the composite matrix function and its first-order/ second- order derivatives. An explicit simple expression is provided for the case of triply repeated eigenvalues.

\subsection{Matrix function}
The matrix function can be obtained by generalizing the scalar-valued function to the matrix-valued function. For example, using series expansion, the exponential matrix function can be defined as
\begin{align}
e^{x}=\sum _{n=0}^{\infty }{\frac {x^{n}}{n!}},\quad
e^{\bm X}:=\sum _{n=0}^{\infty }{\frac {\bm X^{n}}{n!}}.\notag
\end{align}
Matrix function has profound influence on vector-valued differential equations. The generalization of scalar-valued ordinary differential equation to vector-valued ordinary differential equations requires the matrix functions, for example,
\begin{align}
&\frac{d y}{d t}=a y\to y(t)=e^{a t} y(0) \notag\\
&\frac{d \bm y}{d t}=\bm A \bm y \to \bm y(t)=e^{\bm A t} \bm y(0), \notag
\end{align}
where $a,\bm A$ are coefficients, $ \bm y \in \mathbb{C}^{n}, \bm A \in \mathbb{C}^{n \times n}$.
\subsection{Spectral decomposition for matrix function and its derivatives}
In continuum mechanics, the third-order matrix is often used. Let us focus on the symmetric matrix function of third-order
\begin{align}
\bm A=\begin{pmatrix}
a_{11}& a_{12} & a_{13}\\
a_{12}& a_{22} & a_{23}\\
a_{13}& a_{23} & a_{33}\\
\end{pmatrix},
\end{align}
where $a_{ij}:=a_{ij}(b_1,...,b_n)$ are the functions of variables ($b_1,b_2,...,b_n)$.

The eigenvalue decomposition of $\bm A$ is
\begin{align}
\bm A=\sum_{i=1}^3 \lambda_i \bm n_i \otimes \bm n_i,
\end{align}
where $\lambda_i$ and $\bm n_i$ are the distinct eigenvalues and orthogonal unit eigenvectors, respectively. For the case of repeated eigenvalues, the distinct eigenvectors can be achieved by perturbation of the repeated eigenvalues.

Based on spectral decomposition, the matrix function of $\bm A$ can be written as
\begin{align}
f(\bm A)=\sum_{k=1}^3 f(\lambda_k) \bm n_k \otimes \bm n_k=:\begin{pmatrix}f_{11}&f_{12}&f_{13}\\f_{12}&f_{22}&f_{23}\\f_{13}&f_{23}&f_{33}\end{pmatrix}, 
\end{align}
where $f_{ij}$ are the items of the matrix function. 

Using chain-rule, the derivative of $f(\bm A)$ on ($a_{11},a_{12},a_{13},a_{22},a_{23},a_{33}$)
\begin{align}
\pfrac{f(\bm A)}{a_{ij}}=\sum_{k=1}^3 \Big(\pfrac{f(\lambda_k)}{a_{ij}} \bm n_k \otimes \bm n_k+f(\lambda_k) \big(\pfrac{\bm n_k}{a_{ij}}\otimes \bm n_k+ \bm n_k\otimes \pfrac{\bm n_k}{a_{ij}}\big)\Big).\label{eq:dfAda}
\end{align}


When the eigenvalues are identical, the matrix becomes a identity matrix, the matrix derivative has a simple form
\begin{align}
\begin{pmatrix}
\pfrac{f_{11}}{a_{11}}&\pfrac{f_{11}}{a_{12}}&\pfrac{f_{11}}{a_{22}}\\
\pfrac{f_{12}}{a_{11}}&\pfrac{f_{12}}{a_{12}}&\pfrac{f_{12}}{a_{22}}\\
\pfrac{f_{22}}{a_{11}}&\pfrac{f_{22}}{a_{12}}&\pfrac{f_{22}}{a_{22}}\\
\end{pmatrix}=
\begin{pmatrix}
f'(\lambda ) & 0 & 0 \\
0 & f'(\lambda ) & 0 \\
0 & 0 & f'(\lambda ) \\
\end{pmatrix}.
\end{align}

\subsection{Matrix derivative}
Derivatives of eigenvalue and eigenvector with respect to matrix $A$ are
\begin{align}
\partial \lambda_{i}=\bm {n}_{i}^{T} \partial(\bm {A}) \bm {n}_{i}, \quad \partial \bm {n}_{i}=\left(\lambda_{i} \bm {I}-\bm {A}\right)^{+} \partial(\bm {A}) \bm {n}_{i}, \label{eq:derieig}
\end{align}
where $\bm A$ is real and symmetric, $\lambda_{i}$ and $\bm {n}_{i}\left(\bm {n}_{i} \cdot \bm {n}_{j}=\delta_{i j}\right)$ are the distinct eigenvalues and eigenvectors of $\bm A$. The generalized inverse of ($\lambda_{i} \bm {I}-\bm {A}$) has the explicit form: $\left(\lambda_{i} \bm {I}-\bm {A}\right)^{+}=\sum_{j\neq i}^{3} \frac{1}{\lambda_i-\lambda_j} \bm n_j\otimes \bm n_j$. Then the matrix derivative can be written as well as
\begin{align}
\pfrac{ \lambda_{i}}{\bm A}=\bm {n}_{i}\otimes \bm {n}_{i}, \quad \pfrac{ \bm {n}_{i}}{\bm A}=\sum_{j\neq i}^{3} \frac{1}{\lambda_i-\lambda_j} \bm n_j\otimes \bm n_j\otimes \bm {n}_{i} \label{eq:derieig2}
\end{align}

Therefore, the derivative of matrix function $f(\bm A)$ with respect to $b_l$ can be written as
\begin{align}
\pfrac{f(\bm A)}{b_l}=\pfrac{f(\bm A)}{a_{ij}}\pfrac{a_{ij}}{b_l},
\end{align}
where $\pfrac{f(\bm A)}{a_{ij}}$ is computed by Eq.\ref{eq:dfAda}.

For many real applications, it is also required to calculate the second-order derivative of the matrix function with respect to $\bm A$, that is
\begin{align}
&\pfrac{^2 f(\bm A)}{a_{ij}\partial a_{mn}}=\sum_{k=1}^3 \Big(\pfrac{^2f(\lambda_k)}{a_{ij}\partial a_{mn}} \bm n_k \otimes \bm n_k+\pfrac{f(\lambda_k)}{a_{ij}} \pfrac{\bm n_k}{a_{mn}} \otimes \bm n_k\notag\\
&+\pfrac{f(\lambda_k)}{a_{ij}} \bm n_k \otimes \pfrac{\bm n_k}{a_{mn}}+\pfrac{f(\lambda_k)}{a_{mn}} \big(\pfrac{\bm n_k}{a_{ij}}\otimes \bm n_k+ \bm n_k\otimes \pfrac{\bm n_k}{a_{ij}}\big)\notag\\
&+2 f(\lambda_k) \big(\pfrac{^2\bm n_k}{a_{ij}\partial a_{mn}}\otimes \bm n_k+\pfrac{\bm n_k}{a_{ij}}\otimes \pfrac{\bm n_k}{a_{mn}}+ \pfrac{\bm n_k}{a_{mn}}\otimes \pfrac{\bm n_k}{a_{ij}}\notag\\
&+ \bm n_k\otimes \pfrac{^2\bm n_k}{a_{ij}\partial a_{mn}}\big)\Big).
\end{align}
In the above equation, the second order derivative of eigenvalue and eigenvector with respect to the variables are required. The second order derivative can be obtained by calculating the partial derivatives of both sides of Eq.\ref{eq:derieig}, which is written as
\begin{align}
\partial^2 \lambda_{i}&=\partial\bm {n}_{i}^{T} \partial(\bm {A}) \bm {n}_{i}+\bm {n}_{i}^{T} \partial(\bm {A}) \partial\bm {n}_{i}\notag\\
&=(\left(\lambda_{i} \bm {I}-\bm {A}\right)^{+} \partial(\bm {A}) \bm {n}_{i})^{T} \partial(\bm {A}) \bm {n}_{i}+\bm {n}_{i}^{T} \partial(\bm {A}) \left(\lambda_{i} \bm {I}-\bm {A}\right)^{+} \partial(\bm {A}) \bm {n}_{i} \\
\partial^2 \bm {n}_{i}&=\partial\left(\lambda_{i} \bm {I}-\bm {A}\right)^{+} \partial(\bm {A}) \bm {n}_{i}+\left(\lambda_{i} \bm {I}-\bm {A}\right)^{+} \partial(\bm {A}) \partial\bm {n}_{i}\notag\\
&=\partial\left(\lambda_{i} \bm {I}-\bm {A}\right)^{+} \partial(\bm {A}) \bm {n}_{i}+\left(\lambda_{i} \bm {I}-\bm {A}\right)^{+} \partial(\bm {A}) \left(\lambda_{i} \bm {I}-\bm {A}\right)^{+} \partial(\bm {A}) \bm {n}_{i}.
\end{align}
The above two equations contain all terms such as $\pfrac{^2\lambda_k}{a_{ij}\partial a_{mn}}$ and $\pfrac{^2\bm n_k}{a_{ij}\partial a_{mn}}$, which can be extracted by mathematical software such as Mathematica.

Then the second derivative of matrix function $f(\bm A)$ with respect to $b_l$ can be written as
\begin{align}
\pfrac{^2f(\bm A)}{b_l\partial b_k}=\pfrac{^2f(\bm A)}{a_{ij}\partial a_{mn}}\pfrac{a_{mn}}{b_k}\pfrac{a_{ij}}{b_l}+\pfrac{f(\bm A)}{a_{ij}}\pfrac{^2a_{ij}}{b_l\partial b_k}.
\end{align}

\section{Hyperelasticity based on Hencky strain}
Hencky strain is suitable to describe the moderate large deformation of elastic solid \cite{anand1979h}. Hencky strain is described by the logarithmic function of Cauchy-Green tensor. Using conventional small deformation material tensor, the finite strain energy density based on logarithmic strain of left Cauchy-Green tensor ($\bm b=\bm F\bm F^T$) can be written as
\begin{align}
\phi(\bm b_e)=\frac 12\bm \sigma:\bm \varepsilon=\frac{\lambda}{2}\big( \mbox{tr}\bm \varepsilon\big)^2+\mu \bm\varepsilon: \bm\varepsilon,
\end{align}
where $\bm \varepsilon=\frac 12\ln\bm b$ is the logarithmic strain, $\bm \sigma$ is the stress tensor via the conventional 4th-order material tensor $\mathbb C$ as follows.
\begin{align}
\bm {\sigma }=\lambda \mbox {tr} ({\bm {\varepsilon }})\bm {I} +2\mu {\bm {\varepsilon }}={\mathbb {C}}:{\bm {\varepsilon }}\,;\qquad {\mathbb {C}}=\lambda \bm {I} \otimes \bm {I} +2\mu {\bm {I}}.
\end{align}
The variation of strain energy in domain $\Omega$ is derived as
\begin{align}
\delta \Psi&=\int_\Omega\pfrac{\phi}{\bm b}:\delta \bm b=\int_\Omega\pfrac{\phi}{\bm b}:(\bm F\delta \bm F^T+\delta \bm F \bm F^T)\notag\\
&=\int_\Omega 2 \pfrac{\phi}{\bm b}:(\bm F\delta \bm F^T)=\int_\Omega 2\bm F^T \pfrac{\phi}{\bm b}:(\delta \bm F^T)\notag\\
&=\int_\Omega 2 \pfrac{\phi}{\bm b}\bm F:\delta \bm F=-\int_\Omega \nabla\cdot (2 \pfrac{\phi}{\bm b}\bm F)\cdot\delta \bm u
\end{align}
where
\begin{align}
\pfrac{\phi}{\bm b}=\big(\frac 14\mathbb C:\log\bm b\big) :\pfrac{\log\bm b}{\bm b}.
\end{align}
In the derivation, $\pfrac{\phi}{\bm b}$ is symmetric and the relation of $A:(BC)=\mbox{tr}[A^T B C]=B^T A:C=A^T B: C^T$ is used.
In the derivation, the boundary term is neglected for the sake of conciseness. For any $\delta \bm u$, $\delta \Psi=0$ in domain $\Omega$ yields the governing equations
\begin{align}
\nabla \cdot \Big(2 \pfrac{\phi}{\bm b}\bm F\Big)+\bm g=\bm 0,
\end{align}
where $\bm g$ is the body force density.

In order to derive the residual vector and tangent stiffness matrix of the energy functional at a point, the Voigt notation of $\bm b$ is used
\begin{align}
\bar{\bm b}=Voigt(\bm b)=\{b_{11},b_{22},b_{33},b_{23},b_{13},b_{12}\}.
\end{align}
Consider energy functional $\phi:=\phi(\bar{\bm b}(\bm F))$. 
Consider the variation of $\phi$
\begin{align}
\delta \phi&={\pfrac{\phi}{\bar{\bm b}}}\cdot\pfrac{\bar{\bm b}}{\bm F}\cdot \delta \bm F\\
\delta^2 \phi&=(\delta \bm F)^T\cdot\Big((\pfrac{\bar{\bm b}}{\bm F})^T\cdot {\pfrac{^2 \phi}{\bar{\bm b}\partial \bar{\bm b}}}\cdot \pfrac{\bar{\bm b}}{\bm F}+\pfrac{\phi}{\bar{\bm b}}\cdot\pfrac{^2\bar{\bm b}}{\bm F\partial \bm F}\Big)\cdot \delta \bm F,
\end{align}
where
\begin{align}
{R(\bar{\bm {b}})}&={\pfrac{\phi}{\bar{\bm b}}}= \frac 12\bm \sigma:\pfrac{\ln \bm b}{\bar{\bm b}}\\
{K(\bar{\bm b})}&={\pfrac{^2 \phi}{\bar{\bm b}\partial \bar{\bm b}}}= \frac 12\bm \sigma:\pfrac{^2\ln \bm b}{\bar{\bm b}\partial \bar{\bm b}}+ \frac 14\mathbb C:\pfrac{\ln \bm b}{\bar{\bm b}}:\pfrac{\ln \bm b}{\bar{\bm b}}.
\end{align}
The calculation of $\pfrac{\ln \bm b}{\bar{\bm b}}, \pfrac{^2\ln \bm b}{\bar{\bm b}\partial \bar{\bm b}}$ should be done for each sub-term based on the spectral decomposition and be constructed based on the Voigt notation of $\bm b$.

So the residual vector and tangent stiffness matrix at a point can be written as
\begin{align}
\mathbb R&=\pfrac{\phi}{\bar{\bm b}}\pfrac{\bar{\bm b}}{\bm F}\pfrac{\bm F}{\bm U}=R(\bar{\bm b})\pfrac{\bar{\bm b}}{\bm F} \bm B\\
\mathbb K&=(\pfrac{\bm F}{\bm U})^T\Big((\pfrac{\bar{\bm b}}{\bm F})^T\pfrac{^2 \phi}{\bar{\bm b}\partial \bar{\bm b}}\pfrac{\bar{\bm b}}{\bm F}+\pfrac{\phi}{\bar{\bm b}}\pfrac{^2\bar{\bm b}}{\bm F\partial \bm F}\Big)\pfrac{\bm F}{\bm U}\notag\\
&=\bm B^T\Big((\pfrac{\bar{\bm b}}{\bm F})^T K(\bar{\bm b})\pfrac{\bar{\bm b}}{\bm F}+R(\bar{\bm b})\pfrac{^2\bar{\bm b}}{\bm F\partial \bm F}\Big)\bm B,
\end{align}
where $\bm B$ is defined in Eq.\ref{eq:BU3}. In above equations, $\pfrac{\bar{\bm b}}{\bm F}$ and $\pfrac{^2\bar{\bm b}}{\bm F\partial \bm F}$ are independent of the material constitutive. When the residual vector and tangent stiffness matrix are derived and the appropriate boundary conditions are enforced, the Newton iteration scheme can be applied to find the solution.

\section{Variational derivation of finite strain elastoplasticity with phase field model}\label{sec:variJ2phasefield}

\subsection{Multiplicative elastoplasticity}\label{sec:plasmulti}
The solid material is assumed as isotropic elasticity, whose strain energy is expressed using the invariants of the Hencky strain of the form
\begin{align}
\phi(\bm b_e)=\frac 12\bm \sigma:\bm \varepsilon=\frac{\lambda}{2}\big( \mbox{tr}\bm \varepsilon\big)^2+\mu \bm\varepsilon: \bm\varepsilon,
\end{align}
where $\bm b_e$ is the elastic part of the left Lagrange-Green strain tensor.

In multiplicative elastoplasticity theory \cite{simo1992algorithms}, the deformation gradient $\bm F$ can be decomposed into as the elastic part $\bm F_e$ and plastic part $\bm F_p$ as,
\begin{align}
\bm F=\bm F_e \bm F_p.
\end{align}
Then the following standard kinematic quantities (total, elastic, and plastic Cauchy-Green tensors) are
\begin{align}
\bm C=\bm F^T \bm F, \bm C_e=\bm F_e^T \bm F_e, \bm C_p=\bm F_p^T \bm F_p, \bm b_e=\bm F_e\bm F_e^T=\bm F \bm C_p^{-1} \bm F^T. \label{eq:bbe4}
\end{align}
The velocity gradient, its elastic and plastic parts, are defined by
\begin{align}
\bm l=\dot{\bm F}\bm F^{-1}, \bm l_e=\dot{\bm F}_e\bm F_e^{-1}, \bm L_p=\dot{\bm F}_p\bm F_p^{-1}, \bm l_p=\bm l-\bm l_e=\bm F_e \bm L_p \bm F_e^{-1}.
\end{align}

The rate of deformation tensor $\bm d$ and spin tensor $\bm w$ are defined for $\bm l$
\begin{align}
\bm d=\frac 12(\bm l+\bm l^T), \bm w=\frac 12(\bm l-\bm l^T).
\end{align}
Similarly, $\bm l_p$ and $\bm L_p$ can be decomposed into symmetric and antisymmetric parts, namely
\begin{align}
\bm l=\bm d+\bm w, \bm l_p=\bm d_p+\bm w_p, \bm L_p=\bm D_p +\bm W_p.
\end{align}
Based on the above definitions, $\dot{\overline{\bm C_p^{-1}}}$ satisfies the relationship
\begin{align}
\dot{\overline{\bm C_p^{-1}}}=-2 \bm F_p^{-1} \bm D_p \bm F_p^{-T}.\label{eq:dCp1}
\end{align}

The yield function and plastic flow rule can be defined either in the intermediate configuration or in the current configuration. For the former case, the adequate stress is the Mandel stress tensor \cite{Mandel1972PlasticitCE,mandel1973equations}
\begin{align}
\bm M=2\bm C_e \pfrac{\phi_e}{\bm C_e}=\bm F_e^T \bm \tau \bm F_e^{-T},
\end{align}
where $\bm \tau=2\bm b_e \pfrac{\phi_e}{\bm b_e}$ is the Kirchhoff stress. The Mandle stress is symmetric in isotropic elasticity. The yield function of $J_2$ plasticity based on the Mandel stress has the form
\begin{align}
\mathcal F=\sqrt{\frac 32}||\bm M'||-\sigma_y(\alpha),
\end{align}
where $\bm M'=\bm M-1/3 \tr{\bm M} \bm I$, $||\bm A||=\sqrt{\bm A:\bm A}$ and $\sigma_y(\alpha)$ is the yield stress function of hardening variable $\alpha$. $\bm \tau_{dev}=\bm \tau-1/3 \tr{\bm \tau} \bm I$. The associated plastic flow rule,
\begin{align}
\bm D_p=\dot{\gamma} \bm N, \quad \bm N=\pfrac{\phi}{\bm M}=\sqrt{\frac 32}\frac{\bm M'}{||\bm M'||}.
\end{align}
Apparently, $\bm N$ is symmetric.

For the case of yield function in the current configuration, the normal of yield surface $\bm n$ is
\begin{align}
\bm n=\pfrac{\mathcal F}{\bm \tau}=\sqrt{\frac 32}\frac{\bm \tau_{dev}}{||\bm \tau_{dev}||}.
\end{align}
$\bm N$ and $\bm n$ is related by
\begin{align}
\bm N=\bm F_e^{-1} \bm n \bm F_e=\bm{F}_{{e}}^T \bm{n} \bm{F}_{{e}}^{-T}.
\end{align}
Utilizing the transformation rule between $\bm N$ and $\bm n$, Eq.\ref{eq:dCp1} can be rewritten as
\begin{align}
\dot{\overline{\bm C_p^{-1}}}&=-2 \bm F_p^{-1} \bm D_p \bm F_p^{-T}=-2 \dot{\gamma} \bm{F}_p^{-1} \bm{N} \bm{F}_p^{-T}=-2 \dot{\gamma} \bm F_p^{-1}(\bm F_e^{-1} \bm n \bm F_e)(\bm F_p \bm F_p^{-1})\bm F_p^{-T}\notag\\
&=-2 \dot{\gamma}( \bm F_p^{-1}\bm F_e^{-1}) \bm n (\bm F_e\bm F_p)( \bm F_p^{-1}\bm F_p^{-T})
=-2 \dot{\gamma}\bm F^{-1} \bm n \bm F \bm C_p^{-1}.\label{eq:dCp2}
\end{align}

Applying matrix exponential function, the solution of Eq.\ref{eq:dCp2} at step $n+1$ is
\begin{align}
\bm C_{p,n+1}^{-1}=\bm F^{-1}\exp\big(-2(\gamma_{n+1}-\gamma_n)\bm n\big)\bm F \bm C_{p,n}^{-1}.\label{eq:dCp3}
\end{align}
By making use of Eq.\ref{eq:bbe4}(4) $ \bm b_{e,n}=\bm F \bm C_{p,n}^{-1} \bm F^T$, Eq.\ref{eq:dCp3} becomes
\begin{align}
\bm b_{e,n+1}=\exp\big(-2(\gamma_{n+1}-\gamma_n)\bm n\big) \bm b_{e,n}.
\end{align}



\subsection{KKT condition via variational derivation}\label{sec:KKTV}
Many physical problems involving irreversibility can be formulated as the Karush-Kuhn-Tucker (KKT) conditions \cite{karush2014minima,boyd2004convex}.
KKT conditions can be understood as finding the minimization of a functional subject to both equations and inequalities \cite{boyd2004convex}. It generalizes the method of Lagrange multipliers by incorporating inequality constraints. In the following, we will show the KKT conditions can be derived from a functional with constraint variation.

Consider two functions $u(x)$ and $c(x)$, and we restrict that
\begin{align}
\delta u\in (-\epsilon, \epsilon), \delta c \in [0, \epsilon),\label{eq:uc1}
\end{align}
where $\epsilon$ is a small positive number. Eq.\ref{eq:uc1} indicates that $\delta u$ can be varied in any direction and $\delta c$ can be done only in one direction. We may call $\delta c$ is the unidirectional variation.

A functional based on function $u(x)$ and $c(x)$ is conceptually written as
\begin{align}
\Psi(u,c)=\int_\Omega \phi(u,c).\notag
\end{align}
The variation of $\Psi(u,c)$ reads
\begin{align}
\delta \Psi(u,c)=\int_\Omega \pfrac{\phi}{u} \delta u+\pfrac{\phi}{c}\delta c.
\end{align}
Since we constraint that $\delta u\in [-\epsilon, \epsilon], \delta c \in [0, \epsilon]$, it is interesting to see what happens when $\delta \Psi(u,c)=0$ is enforced. $\delta \Psi(u,c)=0$ implies
\begin{align}
\pfrac{\phi}{u} \delta u=0\,\,\forall \delta u\in [-\epsilon, \epsilon],\quad\quad\pfrac{\phi}{c}\delta c=0\,\,\forall \delta c \in [0, \epsilon].
\end{align}
For the first term,
$\pfrac{\phi}{u} \delta u=0, \forall\delta u$ leads to $\pfrac{\phi}{u}=0$, which is the condition of conventional variation. For the second term, the increment of $c$ is only possible when $\pfrac{\phi}{c}=0$. If $\pfrac{\phi}{c}\neq 0$, there is $\delta c=0$. Usually, the sign of $\pfrac{\phi}{c}$ can be determined in advance. If $\pfrac{\phi}{c}|_{c=0}>0$, the second equation is equivalent to
\begin{align}
-\pfrac{\phi}{c}\leq 0,\dot{c}\geq 0, \pfrac{\phi}{c}\dot{c}=0,
\end{align}
which is coincidentally the same as the KKT conditions \cite{karush2014minima} if $\delta c$ is replaced with $\dot{c}$, the time derivative of $c$. In other words, the variation of an energy functional with respect to a variable of non-negative property leads to the KKT conditions.

\subsection{Variational derivation of finite strain J2 plasticity and phase field model}
There are several derivation of plasticity model, the {maximum dissipation rule} \cite{Hackl2007Oct}, the rule of dissipation inequality \cite{simo1992algorithms,borden2016phase}, plastic metric in non-Cartesian coordinate \cite{miehe1998formulation}. The variational derivation presented here offers an alternative in the derivation of finite strain J2 plastic model.

Similar to the phase field model for brittle fracture, we propose the energy functional considering both the elastic energy and plastic energy and using the multiplicative elastoplasticity as
\begin{align}
\Psi(\bm b_e,\bm C_p^{-1},c)&=\int_{\Omega_0} g(c) \big(\phi_+^e(\bm b_e,\bm C_p^{-1})
+\phi^p(\gamma)\big)\notag\\&+\phi_-^e(\bm b_e,\bm C_p^{-1})+G_c (\frac{c^2}{2 \ell}+\frac{\ell}{2} \nabla c\cdot \nabla c) dV,
\end{align}
where $c$ is the phase field with $c=0$ denoting no-damage material and $c=1$ denoting full-damaged material, $g(c)=(1-c)^2$ is the degradation function \cite{miehe2010phase}, $\phi^p(\gamma)$ is the plastic energy depending on the hardening parameter $\gamma$
\begin{align}
\phi^p(\gamma)=\sigma_0 \gamma+\frac 12 h \gamma^2,\label{eq:phip}
\end{align}
where $\sigma_0$ is the initial yield stress. In the absence of phase field, $\bm b_e=\bm F\bm C_p^{-1} \bm F^T,
\bm \tau=2 \bm b_e \pfrac{\phi(\bm b_e)}{\bm b_e}$ as shown in section.\ref{sec:plasmulti}.
$\bm C_p^{-1}$ is the internal variable in the multiplicative elastoplasticity in section \ref{sec:plasmulti}. $\phi_+^e$ and $\phi_-^e$ are the decomposition of $\phi(\bm b_e)$, the stored energy energy per unit volume in the intermediate configuration, given as
\begin{align}
\phi(\bm b_e)=\frac{\lambda}{2}\big( \mbox{tr}\bm\varepsilon\big)^2+\mu \bm\varepsilon: \bm\varepsilon, \mbox{ with } \bm\varepsilon=\frac 12\ln\bm b_e.
\end{align}
The variation of $\Psi$ is
\begin{align}
\delta \Psi=&\int_{\Omega_0} \Big(g' \big(\phi_+^e
+\phi^p\big)\delta c+G_c (\frac{c}{\ell} \delta c+2\ell \nabla c\cdot \nabla \delta c)\Big)\notag\\
&+\big(g \pfrac{\phi_+^e}{\bm b_e}+\pfrac{\phi_-^e}{\bm b_e}\big):\delta \bm b_e+
\pfrac{\phi^p}{\gamma} \delta \gamma.
\end{align}
The second term can be simplified as
\begin{align}
\pfrac{\phi_e}{\bm b_e}:\delta \bm b_e&=\pfrac{\phi_e}{\bm b_e}:\Big(\pfrac{\bm b_e}{\bm F}\delta \bm F+\pfrac{\bm b_e}{\gamma}\delta \gamma\Big)\notag\\
&=\pfrac{\phi_e}{\bm b_e}:\Big(\delta {\bm F}\bm C_p^{-1} \bm F^T+ {\bm F}\bm C_p^{-1} \delta\bm F^T+\pfrac{\bm b_e}{\gamma}\delta \gamma\Big)\notag\\
&=\pfrac{\phi_e}{\bm b_e}:\Big(\delta {\bm F}\bm C_p^{-1} \bm F^T+ {\bm F}\bm C_p^{-1} \delta\bm F^T\Big)+\pfrac{\phi_e}{\bm b_e}:\pfrac{\bm b_e}{\gamma}\delta \gamma\notag\\
&=2\pfrac{\phi_e}{\bm b_e}:\Big( {\bm F}\bm C_p^{-1} \delta\bm F^T\Big)+\pfrac{\phi_e}{\bm b_e}:\pfrac{\bm b_e}{\gamma}\delta \gamma\notag\\
&=2\pfrac{\phi_e}{\bm b_e}:\Big( {\bm F}\bm C_p^{-1} \bm F^T \bm F^{-T}\delta\bm F^T\Big)+\pfrac{\phi_e}{\bm b_e}:\pfrac{\bm b_e}{\gamma}\delta \gamma\notag\\
&=2\pfrac{\phi_e}{\bm b_e}:\Big( \bm b_e \bm F^{-T}\delta\bm F^T\Big)+\pfrac{\phi_e}{\bm b_e}:\pfrac{\bm b_e}{\gamma}\delta \gamma\notag\\
&=(\bm b_e \bm F^{-T})^T 2\pfrac{\phi_e}{\bm b_e}:\delta\bm F^T+\pfrac{\phi_e}{\bm b_e}:\pfrac{\bm b_e}{\gamma}\delta \gamma\notag\\
&= 2\pfrac{\phi_e}{\bm b_e}\bm b_e \bm F^{-T}:\delta\bm F+\pfrac{\phi_e}{\bm b_e}:\pfrac{\bm b_e}{\gamma}\delta \gamma.
\end{align}
Then the variation of $\Psi$ becomes
\begin{align}
&\delta \Psi=\int_{\Omega_0} \Big(g' \big(\phi_+^e
+\phi^p\big)\delta c+G_c (\frac{c}{\ell} \delta c+2\ell \nabla c\cdot \nabla \delta c)\Big)+\notag\\
&2\big(g \pfrac{\phi_+^e}{\bm b_e}+\pfrac{\phi_-^e}{\bm b_e}\big)\bm b_e\bm F^{-T}:\delta \bm F+
\Big(g \big(\pfrac{\phi_+^e}{\bm b_e }:\pfrac{\bm b_e}{\gamma}+\pfrac{\phi^p}{\gamma}\big)+\pfrac{\phi_-^e}{\bm b_e }:\pfrac{\bm b_e}{\gamma}\Big) \delta \gamma.
\end{align}
Denoting $\bm \tau(\bm b_e,c)=2\big(g \pfrac{\phi_+^e}{\bm b_e}+\pfrac{\phi_-^e}{\bm b_e}\big)\bm b_e$ and using
\begin{align}
\pfrac{\bm b_e}{\gamma}=\pfrac{\exp\big(-2(\gamma-\gamma_n)\bm n\big)}{\gamma} \bm b_{e,n}\Big\rvert_{\gamma_n \to \gamma}\approx-2 \bm n \bm b_{e},
\end{align}
the last term can be simplified as
\begin{align}
&\Big(g \big(\pfrac{\phi_+^e}{\bm b_e }:\pfrac{\bm b_e}{\gamma}+\pfrac{\phi^p}{\gamma}\big)+\pfrac{\phi_-^e}{\bm b_e }:\pfrac{\bm b_e}{\gamma}\Big) \delta \gamma\notag\\
&=\Big(g \big(-2\pfrac{\phi_+^e}{\bm b_e } \bm b_{e}:\bm n +\pfrac{\phi^p}{\gamma}\big)-2\pfrac{\phi_-^e}{\bm b_e }\bm b_{e}:\bm n \Big) \delta \gamma\notag\\
&=\Big(g \pfrac{\phi^p}{\gamma}-\bm \tau(\bm b_e,c):\bm n\Big)\delta \gamma\notag\\
&=\Big(g (\sigma_0+h \gamma)-||\bm \tau_{dev}||\Big)\delta \gamma.
\end{align}
Coincidentally, $||\bm \tau_{dev}||=\bm \tau(\bm b_e,c):\bm n$ is the Von-Mises stress and $g (\sigma_0+h \gamma)-||\tau_{dev}||$ is equal to the yield function. It indicates that the yield function is variationally consistent.

Then the variation of $\Psi$ becomes
\begin{align}
&\delta \Psi=\int_{\Omega_0} \Big(g' \big(\phi_+^e
+\phi^p\big)\delta c+G_c (\frac{c}{\ell} \delta c+\ell \nabla c\cdot \nabla \delta c)\Big)+\notag\\
&\bm \tau(\bm b_e,c) \bm F^{-T}:\delta \bm F+
\Big(g (\sigma_0+h \gamma)-||\bm \tau_{dev}||)\delta \gamma.
\end{align}
Using integration by parts and for any $\delta c, \delta \bm u$, $\delta \Psi=0$ yields the governing equations
\begin{subequations}
\begin{align}
\nabla \cdot (\bm \tau(\bm b_e,c) \bm F^{-T})+\bm g=0\\
g' \big(\phi_+^e
+\phi^p\big)+G_c (\frac{c}{\ell} -\ell \nabla^2 c)=0.
\end{align}
\end{subequations}

Consider the fact that the $g (\sigma_0+h \gamma)-||\bm \tau_{dev}||$ is positive when the von-Mises stress is small and $\delta \gamma$ represents the plastic increment which is always non-negative, $\delta \Psi=0$ requires its last term being zero. The only possibility is that
\begin{align}
||\bm \tau_{dev}||-g (\sigma_0+h \gamma) \leq 0, \dot{\gamma}\geq 0,\big(||\bm \tau_{dev}||-g (\sigma_0+h \gamma) \big)\cdot \dot{\gamma}=0,
\end{align}
which yields Karush–Kuhn–Tucker (KKT) conditions. The derivation is similar to that in section \ref{sec:KKTV}.

Two typical decompositions of the elastic energy either based on the eigenvalue decomposition of strain tensor
\begin{align}
\phi^e_{\pm}=\frac 12 \lambda \langle \varepsilon_v^e\rangle_{\pm}^2+\mu (\langle \varepsilon_1^e\rangle_{\pm}^2+\langle \varepsilon_2^e\rangle_{\pm}^2+\langle \varepsilon_3^e\rangle_{\pm}^2),
\end{align}
or the volumetric-deviatoric decomposition of the strain tensor
\begin{subequations}
\begin{align}
\phi^e_{+}&=\frac 12 (\lambda+\frac{2\mu}{3}) \langle \mbox{tr} \bm\varepsilon^e\rangle_{+}^2+\mu \bm\varepsilon^e_{dev}:\bm\varepsilon^e_{dev}\\
\phi^e_{-}&=\frac 12 (\lambda+\frac{2\mu}{3}) \langle \mbox{tr} \bm\varepsilon^e\rangle_{-}^2,
\end{align}
\end{subequations}
where $ \bm\varepsilon^e_{dev}=\bm\varepsilon^e-\frac 13 \mbox{tr} \bm\varepsilon^e \bm I$.

In the current paper, considering the fact that the J2 finite strain plasticity depends on the deviatoric part of the strain tensor only, we employ the second version of the energy decomposition of strain energy density. In addition, the plasticity deformation contributes to the fracture process, the plastic work Eq.\ref{eq:phip} is added to the crack driving energy.
%
%

When large strain elastoplasticity is considered, the ratio of plastic energy is dominant compared to the elastic energy. The damage increases significantly when the plastic energy approaches the threshold value. When only the contribution of $\phi_+^e$ is considered, the fracture is brittle type, otherwise ductile type.

The damage affects both the elastic and plastic deformation. When localization occurs, undegraded stress at the crack tips increases further while stress on other parts undergo an unloading condition. So the effective Von-Mises stress depends on the damage status of that point, and the yield function is modified as
\begin{align}
\mathcal F=g(s) \sigma_{vm}-\sigma_y(\gamma) ,\label{eq:Fgs}
\end{align}
where the Von-Mises stress is
\begin{align}
\bm \tau_{dev}=\bm \tau-\frac 13 \mbox{tr}(\bm\tau) \bm I,
\sigma_{vm}=\sqrt{\frac 23 \bm \tau_{dev}:\bm \tau_{dev}}.
\end{align}
\section{Numerical implementation}
We employ the NOM scheme outlined in Section \ref{sec:nommethod} to express the gradient of the displacement field. The algorithm based on spectral decomposition presented in Section \ref{sec:matrixFun} is used to derive and calculate the exponential matrix function, logarithmic matrix function and their first-order/second-order partial derivatives. The convergence of the numerical solution requires the local convergence of the J2 plastic model, the global convergence of the phase field and the displacement field. For the simplicity of numerical implementation, the staggered scheme in phase field model is used. The phase field model is updated using the current positive energy and historic energy as the crack driving state function. In the mechanical field, the phase field is used as a material parameter to modify the stress and the associated plastic model. Two models are calculated repeatedly until both the variations of phase field and displacement fall inside the tolerance. In the current work, we select $\epsilon_{tol}=10^{-6}$.

The J2 plastic model with phase field is achieved by using the degradation function of the von-Mises stress, see Eq.\ref{eq:Fgs}. The actual stress is used to determine whether the yield criterion is reached. By doing so, there are two benefits: a) the actual stress located in the damaged zone is small thus avoiding the local iteration of plasticity; b) both the localizations of the damage field and plastic zone are located in a small zone. If the yield stress is degraded in the same manner, the local iteration of plasticity is  required as long as the strain increases even if the material is severe damaged.

\subsection{Newton-Iteration at one material point}

The finite strain J2 plasticity model requires Newton iteration method to find the solution. According Section \ref{sec:variJ2phasefield}, the local solution of J2 plastity is equivalent to  find $\bm C_p^{-1},\gamma$ at current step based on trial deformation gradient $\bm F$ and internal variables $\bm C_{p,n}^{-1}, \gamma_n$ at previous step. The main formulas are put together as 
\begin{align}
&\bm b=\bm F\bm F^T \to \bm b_e=\bm F \bm C_p^{-1}\bm F^T\notag\\
&\phi=\hat{\phi}(\frac 12\ln \bm b_e),\,\bm \tau=2 \bm b_e\pfrac{\phi}{\bm b_e}, \bm \tau_{dev}=\bm \tau-\frac 13 (\tr{\bm \tau}) \bm I\notag\\
&\mathcal F=(\frac 32 \bm \tau_{dev}\cdot \bm \tau_{dev})^{\frac 12}-\sigma_y(\gamma),\,{\bm n}=\pfrac{\mathcal F}{\bm \tau}=\sqrt{\frac 32} \frac{\bm \tau_{dev}}{\|\bm \tau_{dev}\|}\notag\\
&\bm Z=\bm F\bm C_p^{-1}-\exp(-2 (\gamma-\gamma_n){\bm n}) \bm F\bm C_{p,n}^{-1}\notag\\
&\bm h=\{C_{p,11}^{-1},C_{p,22}^{-1},C_{p,33}^{-1},C_{p,23}^{-1},C_{p,13}^{-1},C_{p,12}^{-1},\gamma\}\notag\\
&\bm Q=\{Z_{11},Z_{22},Z_{33},Z_{23},Z_{13},Z_{12},\mathcal F\}.\notag
\end{align}
Vector $\bm Q$ is a nonlinear vector-valued function with unknowns specified by $\bm h$. The Newton-Raphson algorithm can be used to find the solution of $\bm Q(\bm h)=0$ by
\begin{align}
\bm h_{k+1}=\bm h_{k}-\Big(\pfrac{\bm Q(\bm h_k)}{\bm h_k}\Big)^{-1} \bm Q(\bm h_k),\notag
\end{align}
where $\bm h_{k+1}$ is the vector value at iteration step $k+1$ and $\pfrac{\bm Q(\bm h_k)}{\bm h_k},(\pfrac{\bm Q(\bm h_k)}{\bm h_k})^{-1}$ are given by
\begin{align}
\pfrac{\bm Q}{\bm h}=\begin{pmatrix}\pfrac{Z_{11}}{C_{11}^{p-1}}& \pfrac{Z_{11}}{C_{22}^{p-1}}&\dots& \pfrac{Z_{11}}{\gamma}\\
\pfrac{Z_{22}}{C_{11}^{p-1}}& \pfrac{Z_{22}}{C_{22}^{p-1}}&\dots& \pfrac{Z_{22}}{\gamma}\\
\vdots&\vdots & \ddots & \vdots\\
\pfrac{\mathcal F}{C_{11}^{p-1}}& \pfrac{\mathcal F}{C_{22}^{p-1}}&\dots& \pfrac{\mathcal F}{\gamma}\\
\end{pmatrix}
,\,(\pfrac{\bm Q}{\bm h})^{-1}=\begin{pmatrix}\pfrac{C_{11}^{p-1}}{Z_{11}}& \pfrac{C_{11}^{p-1}}{Z_{22}} &\dots&\pfrac{C_{11}^{p-1}}{\mathcal F} \\
\pfrac{C_{22}^{p-1}}{Z_{11}}&\pfrac{C_{22}^{p-1}}{Z_{22}}&\dots&\pfrac{C_{22}^{p-1}}{\mathcal F} \\
\vdots&\vdots & \ddots & \vdots\\
\pfrac{\gamma}{Z_{11}}&\pfrac{\gamma}{Z_{22}} &\dots& \pfrac{\gamma}{\mathcal F}\\
\end{pmatrix}.
\end{align}

In the converged iteration of plasticity, $\bm Q=0$ is a function of $\bm F$ and $\bm h$. Differentiating on $\bm Q=0$, we have
\begin{align}
\pfrac{\bm Q}{\bar{\bm F}}+\pfrac{\bm Q}{\bm h}\pfrac{\bm h}{\bar{\bm F}}=0 \to \pfrac{\bm h}{\bar{\bm F}}=-(\pfrac{\bm Q}{\bm h})^{-1}\pfrac{\bm Q}{\bar{\bm F}}.
\end{align}
In above formula, the involved matrix expressions include $ \pfrac{\bm h}{\bm {\bar{F}}},\pfrac{\bm Q}{\bar{\bm F}}$, which are given as follows.
\begin{align}
\pfrac{\bm h}{\bm {\bar{F}}}=\begin{pmatrix}\pfrac{C_{11}^{p-1}}{F_{11}}& \pfrac{C_{11}^{p-1}}{F_{12}}&\dots& \pfrac{C_{11}^{p-1}}{F_{33}}\\
\pfrac{C_{22}^{p-1}}{F_{11}}& \pfrac{C_{22}^{p-1}}{F_{12}}&\dots& \pfrac{C_{22}^{p-1}}{F_{33}}\\
\vdots&\vdots & \ddots & \vdots\\
\pfrac{\gamma}{F_{11}}& \pfrac{\gamma}{F_{12}}&\dots& \pfrac{\gamma}{F_{33}}\\
\end{pmatrix}
,\quad\pfrac{\bm Q}{\bar{\bm F}}=\begin{pmatrix}\pfrac{Z_{11}}{F_{11}}& \pfrac{Z_{11}}{F_{12}}&\dots& \pfrac{Z_{11}}{F_{33}}\\
\pfrac{Z_{22}}{F_{11}}& \pfrac{Z_{22}}{F_{12}}&\dots& \pfrac{Z_{22}}{F_{33}}\\
\vdots&\vdots & \ddots & \vdots\\
\pfrac{\mathcal F}{F_{11}}& \pfrac{\mathcal F}{F_{12}}&\dots& \pfrac{\mathcal F}{F_{33}}\\
\end{pmatrix}.
\end{align}
In sum, the algorithm to implement the local iteration and consistent tangent stiffness matrix are summarized in \textbf{Algorithm 1}.

\quad\fbox{\begin{minipage}{20em}
\textbf{Algorithm 1:} Local iteration for J2 plasticity
\vspace{-1.4cm}
\begin{align}
&\mathcal F^{trial}=\mathcal F(\bm F,\bm h_{n})\notag\\[-10pt]
&\mbox{IF }\mathcal F^{trial}<0, \mbox{ then in elastic state}\notag\\[-10pt]
&\quad \bm h=\bm h_{n}\notag\\[-10pt]
&\mbox{ELSE: in plastic state, begin local plastic iteration}\notag\\[-10pt]
&\quad \bm h=\bm h_{n}\notag\\[-10pt]
&\quad \mbox{ REPEAT}\notag\\[-6pt]
&\quad\quad \bm A=\pfrac{\bm Q(\bm F,\bm h,\bm h_{n})}{\bm h}\notag\\[-6pt]
&\quad\quad \Delta \bm h=-\bm A^{-1} \bm Q\notag\\[-10pt]
&\quad\quad \bm h=\bm h+\Delta \bm h\notag\\[-10pt]
&\quad \mbox{ UNTIL }\|\Delta \bm h\|\leq \varepsilon_h\notag\\[-5pt]
&\quad \bm h=\bm h, \pfrac{\bm h}{\bm F}=-\bm A(\bm F,\bm h,\bm h_{,n})^{-1} \pfrac{\bm Q(\bm F,\bm h,\bm h_{,n})}{\bm F}\notag\\[-10pt]
&\mbox{END IF}\notag
\end{align}
\end{minipage}}

Let $\bm F$ be written in vector form $\bar{\bm F}$ as
\begin{align}
\bar{\bm F}=\{F_{11},F_{12},F_{13},F_{21},F_{22},F_{23},F_{31},F_{32},F_{33}\}.
\end{align}
Let the list of unknown displacement in support denoted by $\bm U$, $\bar{\bm F}$ is a function of $\bm U$.  According to Eq.\ref{eq:BU3}, we have $\pfrac{\bar{\bm F}}{\bm U}=\bm B$. 
 
Then the local residual vector and consistent tangent stiffness matrix of one particle based on unknown vector $\bm U$ are calculated by
\begin{align}
\mathbb R=\Delta V\pfrac{\phi_e}{\bar{\bm F}}\pfrac{\bar{\bm F}}{\bm U}=\Delta V\pfrac{\phi_e}{\bar{\bm F}}\bm B,
\end{align}
\begin{align}
\mathbb K&=\pfrac{\mathbb R}{\bm U}=\Delta V\Big(\pfrac{\bar{\bm F}}{\bm U}\Big)^T\big(\pfrac{^2\phi_e}{\bar{\bm F}\partial\bar{\bm F}}+\pfrac{^2\phi_e}{\bar{\bm F}\partial\bm h}\pfrac{\bm h}{\bar{\bm F}}\big)\pfrac{\bar{\bm F}}{\bm U}\notag\\
&=\Delta V\bm B^T\Big(\pfrac{^2\phi_e}{\bar{\bm F}\partial\bar{\bm F}}-\pfrac{^2\phi_e}{\bar{\bm F}\partial\bm h}(\pfrac{\bm Q}{\bm h})^{-1}\pfrac{\bm Q}{\bar{\bm F}}\Big)\bm B,
\end{align}
where $\Delta V$ is the volume of the material point.
%
%
%
%
After assembling $\mathbb R_i$ and $\mathbb K_i$ of all particles into the global residual and global stiffness matrix, the standard Newton-Iteration algorithm can be applied.

The numerical method in the paper is implemented using the nonlocal operator method. The number of particles in support is chosen as 9 in 2D and 27 in 3D, where the support size is determined by the furthest particles in support for each particle. The weight function is selected as $w(r)=1/r^2$. The nonlocal gradient in discrete form plays the same role as the shape function in finite element method. Since nodal integral is used, the hourglass energy functional is added to the weak form to suppress the zero-energy mode. The calculation of residual and tangent stiffness matrix for hourglass energy function is based on Eq.\ref{eq:khg3}. 

\section{Numerical examples}

%
%
%
%

\subsection{Phase field based on Hencky strain without plasticity: Single-edge-notched tension test}
The current phase field scheme can recover the small strain phase field model when the $G_c$ in small strain brittle fracture is used. The plasticity deformation can be prevented by making the yield stress large enough. In this subsection, we model the single-edge-notched tension test, which is a squared plate with initial notched crack as shown in Fig.\ref{fig:notchPlate}. The material parameters are set as $\lambda=121.1538$ kN/mm$^2$ and $\mu=80.7692$ kN/mm$^2$ for elastic constants, $G_c=2.7\times 10^{-3}$ kN/mm for critical energy release rate. These parameters are identical to that used in the small strain brittle fracture phase field in Ref \cite{miehe2010thermodynamically}. Two displacement conditions are tested: Case a) for tensile boundary condition and Case b) for shear boundary conditions. The plate is discretized into $100\times 100$ material points, the phase field length scale is selected as $l=10^{-2}$ mm. The displacement load is monotonic applied with fixed displacement increment $\Delta u=1\times 10^{-5}$ mm.

\begin{figure}
	\centering
		\includegraphics[width=6cm]{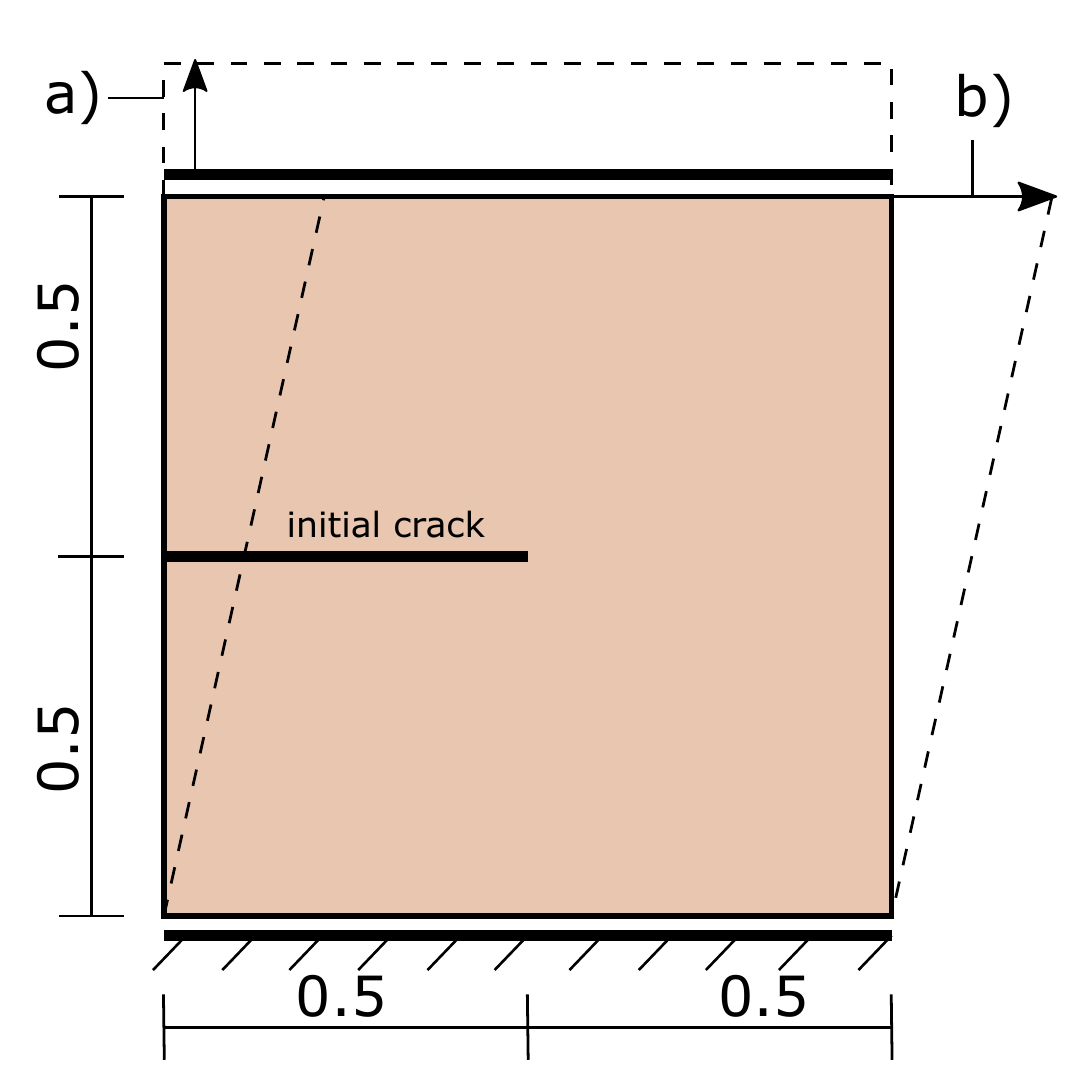}
	\caption{Single-edge-notched tension test. Geometry and Case a for tensile boundary condition and Case b for shear boundary condition.}
	\label{fig:notchPlate}
\end{figure}

\begin{figure}
	\centering
		\includegraphics[width=7cm]{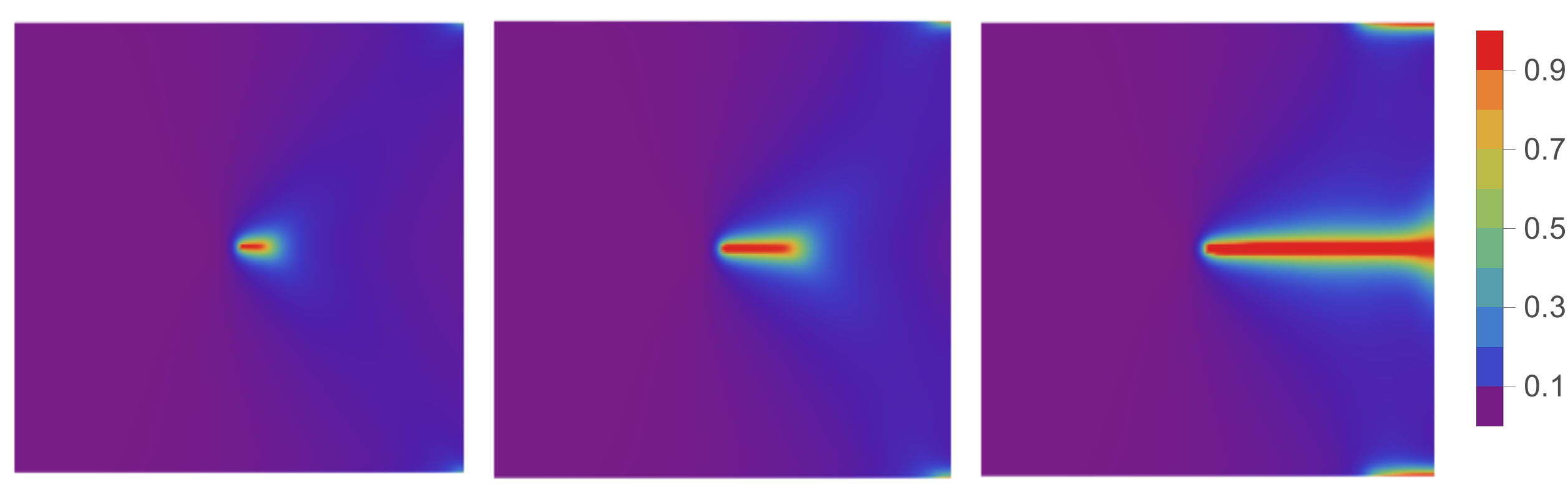}
	\caption{Phase field of single-edge-notched tensile test for brittle fracture.}
	\label{fig:notchplateTensionPf}
\end{figure}

\begin{figure}
	\centering
		\includegraphics[width=4cm]{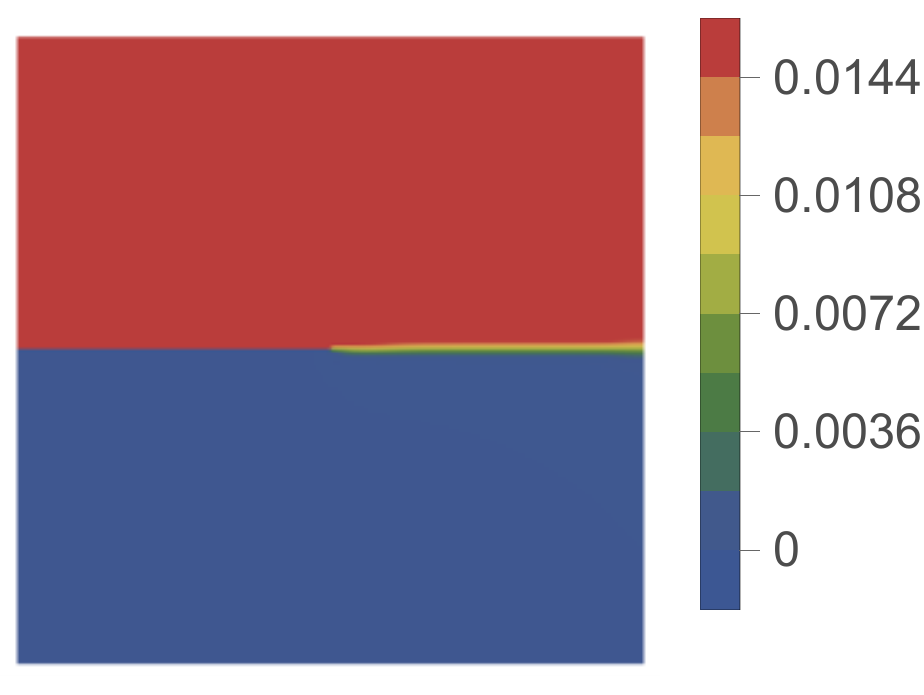}
	\caption{Phase field of single-edge-notched tensile test for brittle fracture.}
	\label{fig:notchplateTensionPf}
\end{figure}
\begin{figure}
	\centering
		\includegraphics[width=8cm]{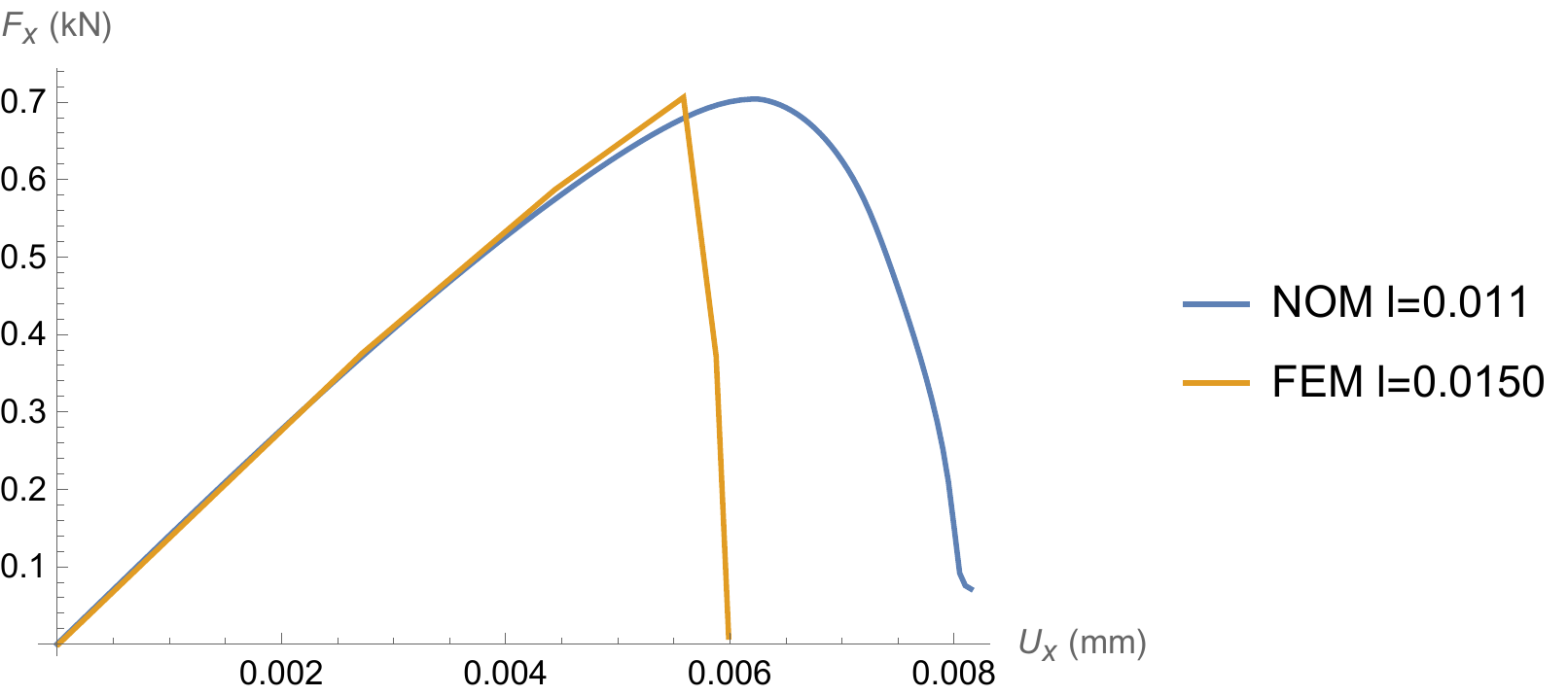}
	\caption{Displacement-load curve of single-edge-notched tensile test for brittle fracture. FEM results come from \cite{miehe2010phase}.}
	\label{fig:notchPlateTensileLoad}
\end{figure}

The evolution of the phase field for tensile tests with a discretization of $200\times 200$ is depicted in Fig.\ref{fig:notchplateTensionPf}. The final displacement field in y-direction is given in Fig.\ref{fig:notchplateTensionPf}. The displacement-load curve is plotted in Fig.\ref{fig:notchPlateTensileLoad}.

\begin{figure}
	\centering
		\includegraphics[width=10cm]{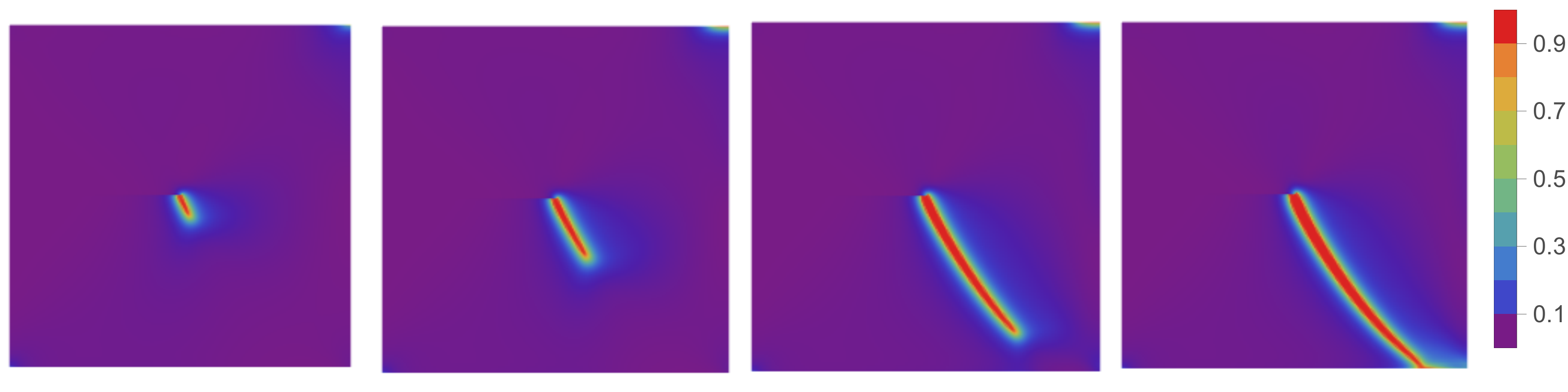}
	\caption{Phase field of single-edge-notched shear test for brittle fracture at $u_x=(0.013, 0.015, 0.022, 0.03)mm$.}
	\label{fig:notchPlateshearphasefield}
\end{figure}

\begin{figure}
	\centering
		\includegraphics[width=7cm]{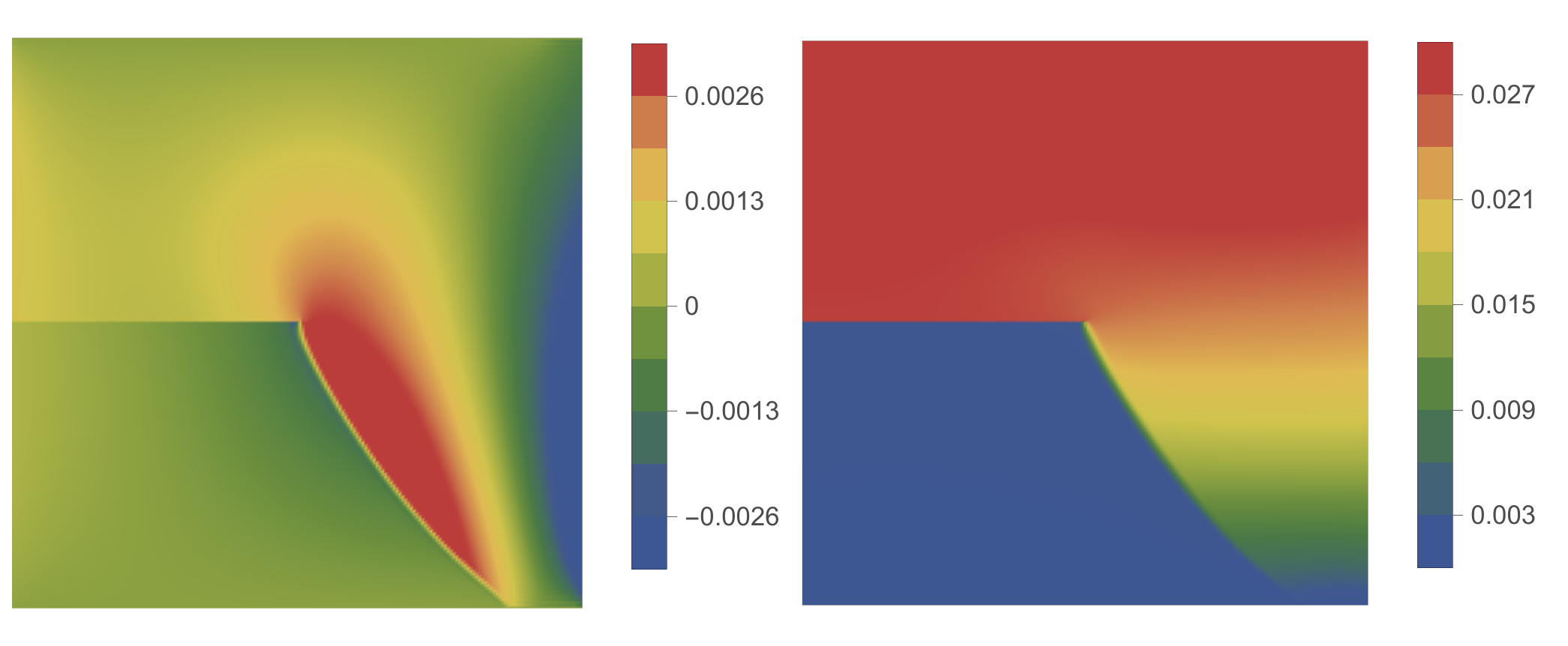}
	\caption{Displacement of single-edge-notched shear test for brittle fracture.}
	\label{fig:notchplateShearElastic4e4ux}
\end{figure}
The phase field evolution of shear test for brittle fracture is depicted in Fig.\ref{fig:notchPlateshearphasefield}. The displacement field is given in Fig.\ref{fig:notchplateShearElastic4e4ux}. When plasticity is not involved and the critical energy release rate for brittle fracture is used, the Hencky strain phase field model can recover the small strain version. The load curve of notched plate modeled by the current scheme is compared to the traditional small strain finite element method in Fig.\ref{fig:notchPlateShearLoad}. The phase field length scale dependence is observed in the current simulation. On the other hand, the Hencky strain formulation considers the geometric nonlinearity and shows a slightly smooth degradation.

\begin{figure}
	\centering
		\includegraphics[width=8cm]{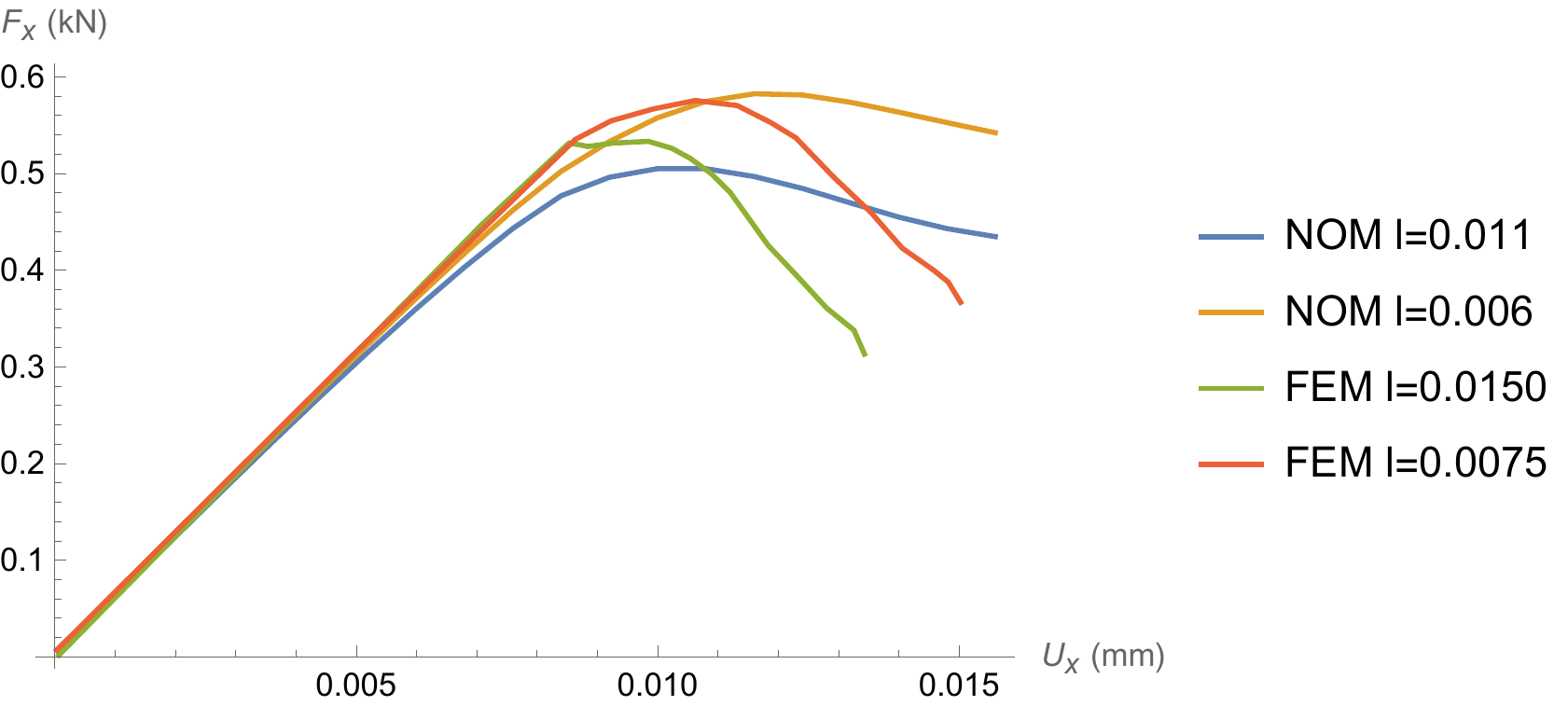}
	\caption{Displacement-load curve of single-edge-notched shear test for brittle fracture. FEM results come from \cite{miehe2010phase}.}
	\label{fig:notchPlateShearLoad}
\end{figure}

\subsection{Phase field based on Hencky strain elastoplasticity: Single-edge-notched test with ductile fracture}

%

The plate is discretized into 200x200 particles. The ductile critical energy release rate is selected as $G_c^d=(5 G_c,10 G_c,15 G_c)$, where $G_c=2.7\times 10^{-3}$ kN/mm is the critical energy release for brittle fracture. The Hencky strain J2 plasticity is used. For the case of $G_c^d=5 G_c$, the final phase field, displacement field, equivalent plasticity and deformed configuration are provided in Fig.\ref{fig:notchplateTensileDuctile4e4}. It can be observed that the crack propagates along the direction of the initial crack and the plastic deformation and ductile fracture happens in the same region. The plastic energy drives the propagation of the phase field dominantly. Conversely, the evolution of the phase field unloads the internal force in other regions, which induces a strain localization and large plastic deformation only in the damaged region. Compared with the brittle phase field, the notched plate undergoes a large deformation before the phase field becomes significant. The reaction forces for different $G_c^d$ is provided in Fig.\ref{fig:plateDuctileGcds}. The displacement-load decreases gradually due to the dissipation in plastic deformation.

\begin{figure}[htp]
	\centering
		\includegraphics[width=10cm]{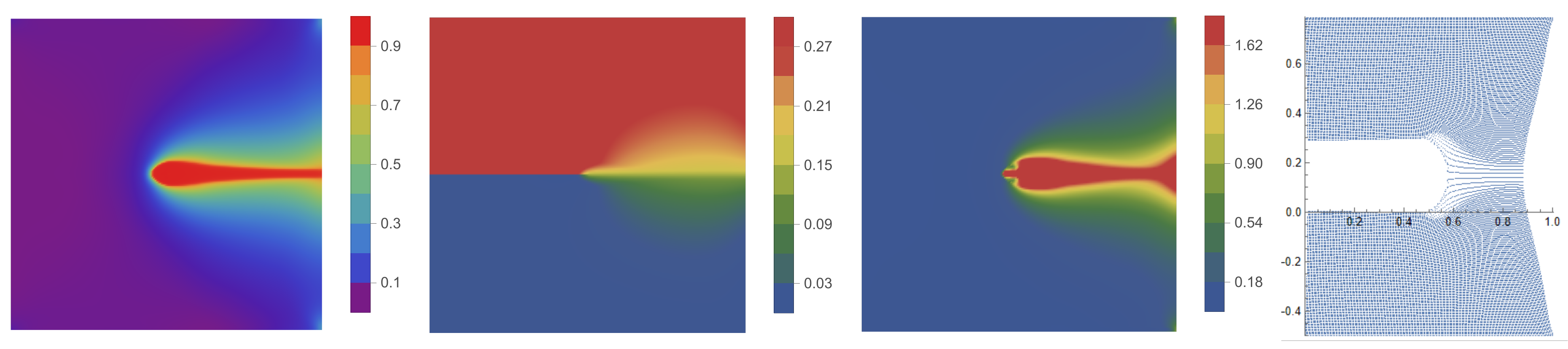}
	\caption{Single-edge-notched tension test a) phase field, b) displacement in y-direction, c) equivalent plasticity and d) deformed configuration.}
	\label{fig:notchplateTensileDuctile4e4}
\end{figure}
\begin{figure}
	\centering
		\includegraphics[width=7cm]{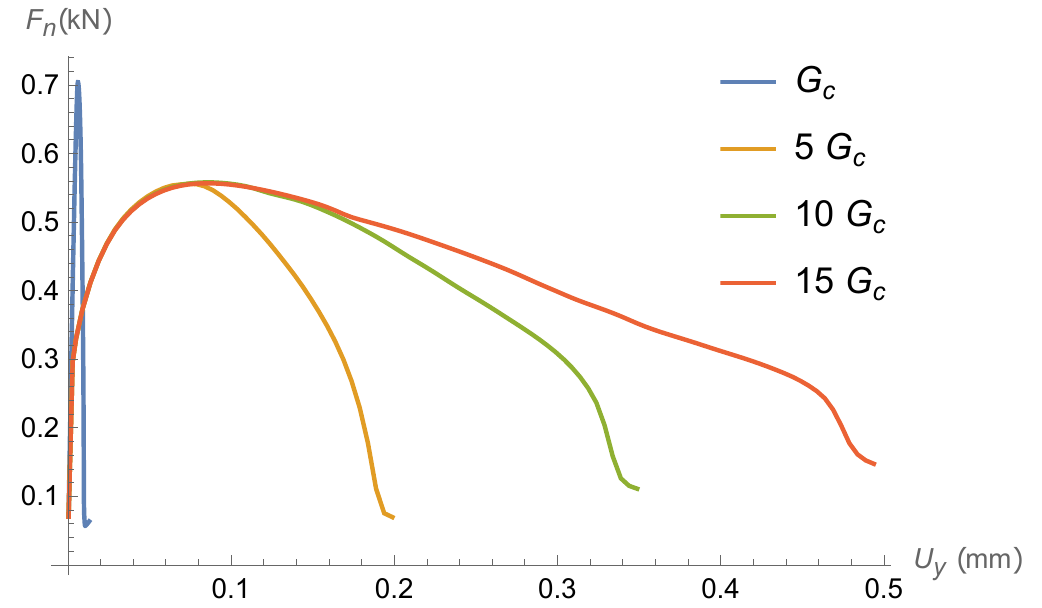}
	\caption{Load curves of single-edge-notched tension test. The curve of $G_c$ is for brittle fracture.}
	\label{fig:plateDuctileGcds}
\end{figure}


\begin{figure}[htp]
	\centering
		\includegraphics[width=10cm]{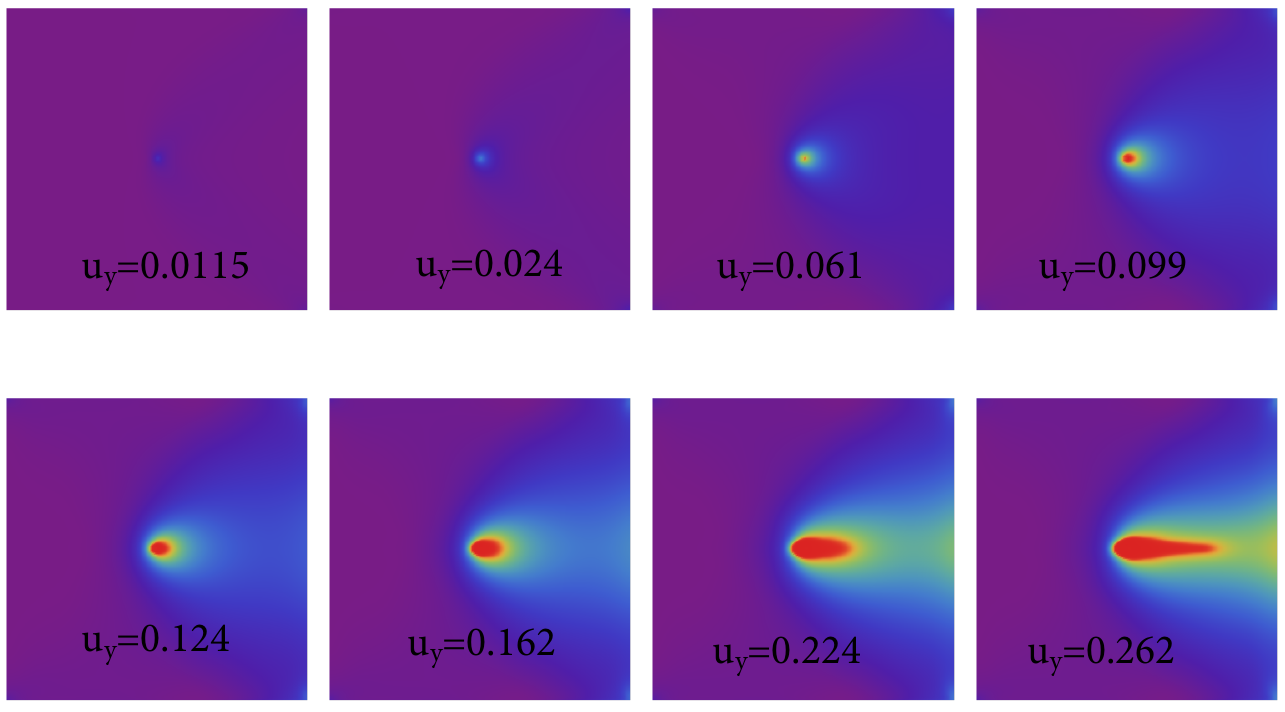}
	\caption{Single-edge-notched tension test: evolution of phase field.}
	\label{fig:NotchTensilePhaseFieldEvolution}
\end{figure}
\begin{figure}[htp]
	\centering
		\includegraphics[width=10cm]{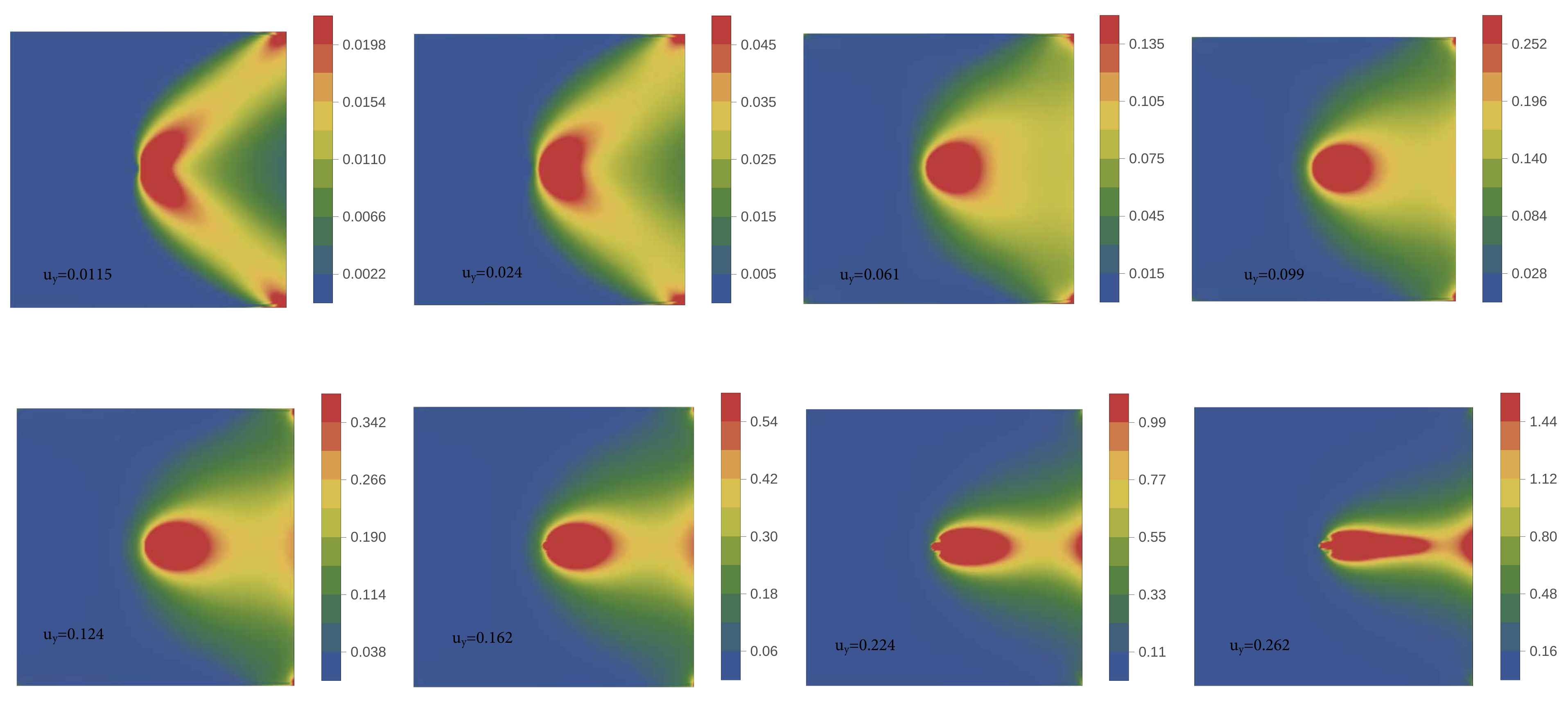}
	\caption{Single-edge-notched tension test: evolution of equivalent plasticity.}
	\label{fig:NotchTensilePlasticityEvolution}
\end{figure}

The evolution of phase field for different time instants is illustrated in Fig.\ref{fig:NotchTensilePhaseFieldEvolution} and the equivalent plasticity field is shown in Fig.\ref{fig:NotchTensilePlasticityEvolution}. The tensile load results in a large deformation along the path of phase field region. It is observed that the large plastic region and the phase field region coincide, which shows that the energy due to plastic deformation contributes significantly to the development of phase field.

\subsection{Isotropic hardening J2 plasticity: necking of a rod}\label{subsec:J2rodnecking}
Necking of a rod is a classical benchmark problem for isotropic elastic-plastic material subjected to finite plasticity, see for example \cite{simo1992algorithms,papadopoulos1998general, miehe2002anisotropic}. The length of the rod in the initial configuration is $l=53.34$ mm, the radius $r_0=6.4135$ mm. A finer discretization is employed in the middle of the rod close to the necked zone. In order to trigger a necking, an imperfection of the center of the rod is introduced by reducing the radius to $r=0.982 r_0$ gradually. The nonlinear isotropic hardening response is described by the yielding function
\begin{align}
\sigma_y=\sigma_0+ h \gamma+(y_{\infty}-y_0) (1-\mbox{exp}(-\Delta \gamma)),
\end{align}
whose material parameters are given in Table \ref{tab:rodnecking}.

\begin{table}[h]
\begin{center}
\begin{tabular}{ l l }
\hline
Material parameters &\\ \hline
Bulk modulus & $\kappa=164.206$ kN/mm$^2$\\
Shear modulus & $\mu=80.1938$ kN/mm$^2$ \\
Initial yield stress & $y_0=0.45$ kN/mm$^2$\\
Infinite yield stress & $y_{\infty}=0.715 $ kN/mm$^2$\\
Hardening modulus& $h=0.12924$ kN/mm$^2$\\
Saturation parameter & $\Delta=16.93$\\\hline
\end{tabular}
\caption{Necking of a rod}\label{tab:rodnecking}
\end{center}
\end{table}
\begin{figure}
	\centering
		\includegraphics[width=5cm]{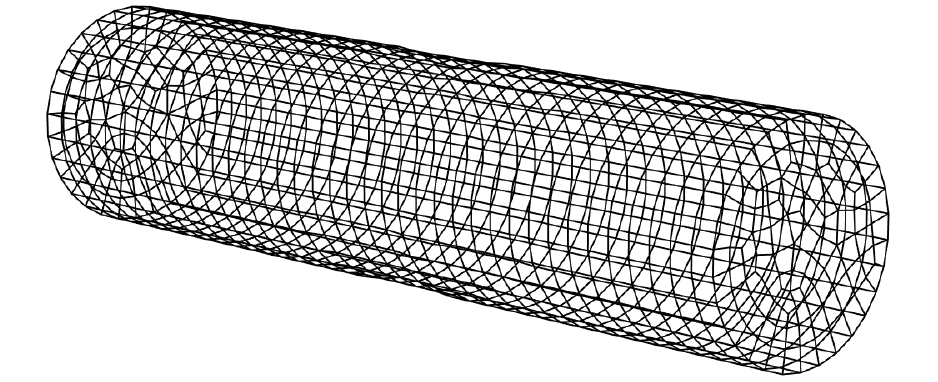}
	\caption{Bar necking: mesh with $\approx 5k$ nodes.}
	\label{fig:meshNeckbar}
\end{figure}
The number of nodes is 4758 ($\approx 5k$) for coarse mesh as shown in Fig.\ref{fig:meshNeckbar} and 49320 ($\approx 50k$) for refined mesh.

\begin{figure}
\centering
\subfigure[$U_z$]{
\label{fig:4kdisp}
\includegraphics[width=.2\textwidth]{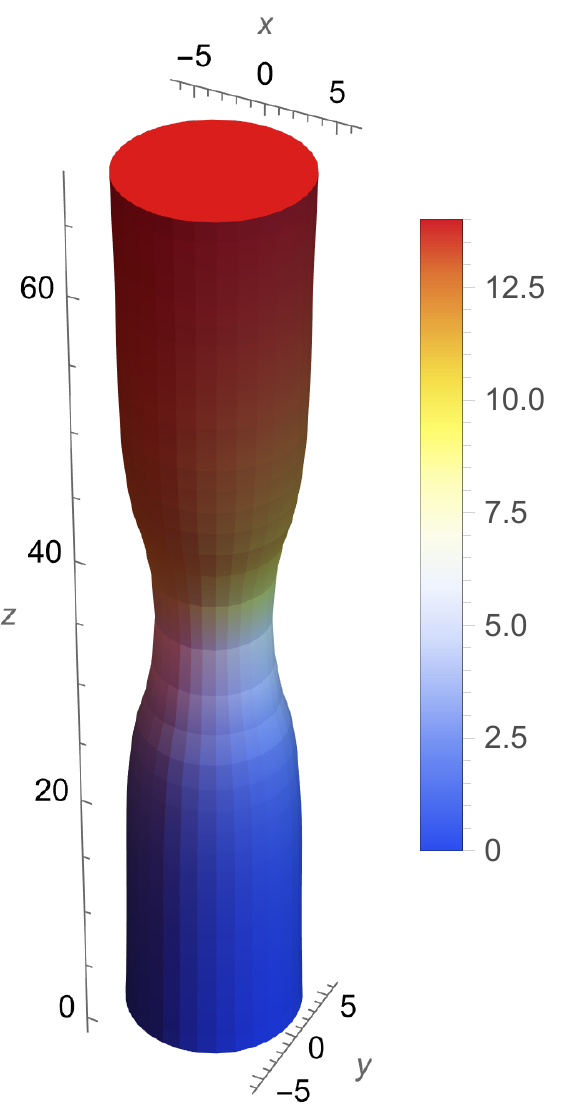}}
\subfigure[$\gamma$]{
\label{fig:4kplastic}
\includegraphics[width=.2\textwidth]{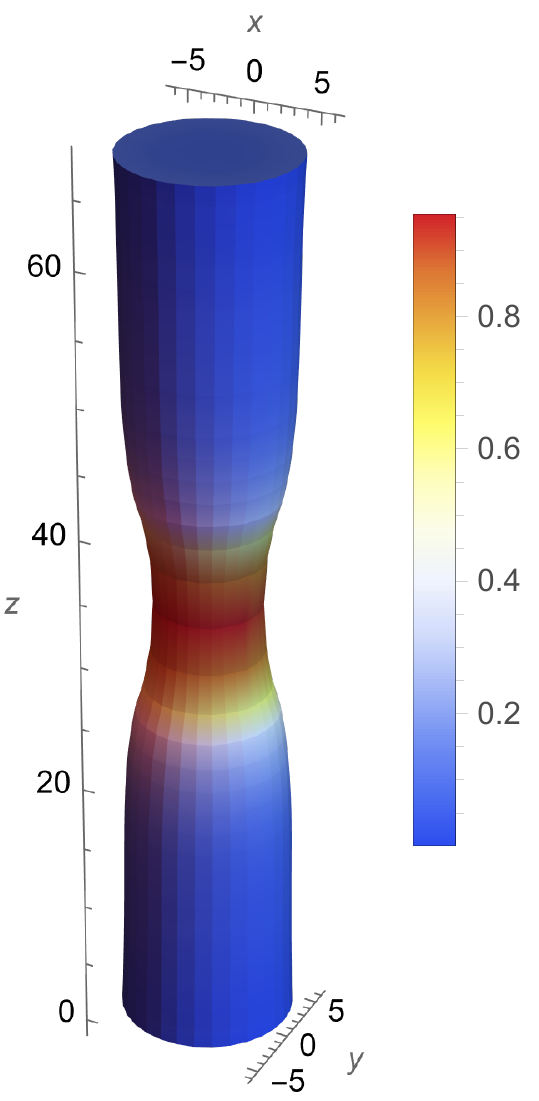}}
\subfigure[$U_z$]{
\label{fig:40kdisp}
\includegraphics[width=.2\textwidth]{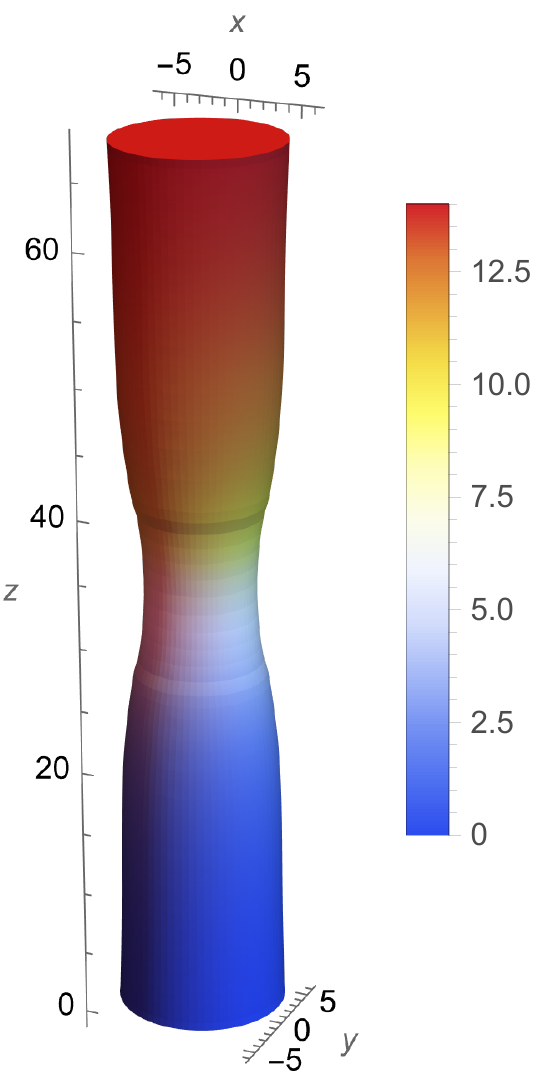}}
\subfigure[$\gamma$]{
\label{fig:40kplastic}
\includegraphics[width=.2\textwidth]{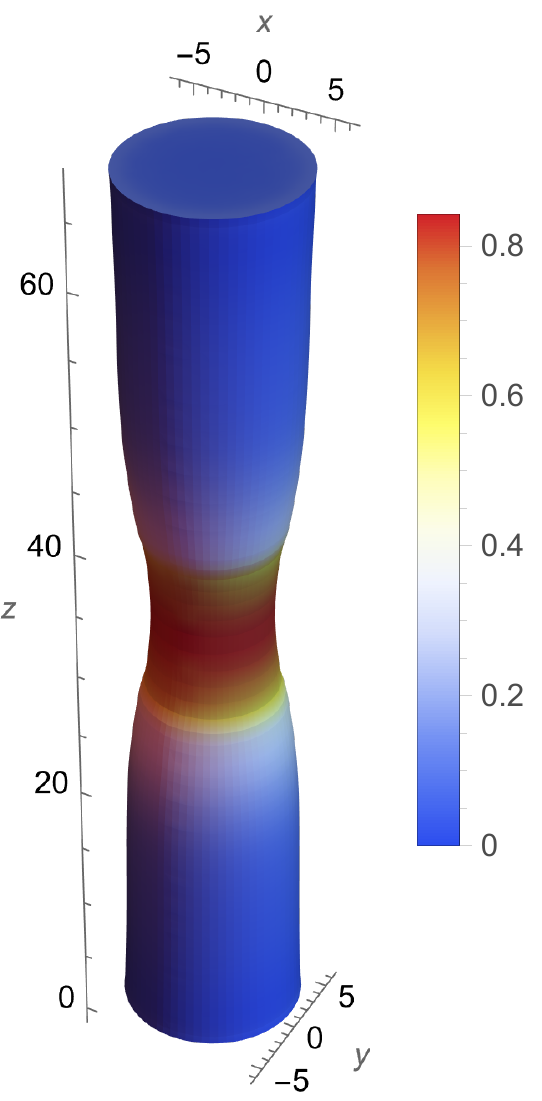}}\\
\caption{Deformed configuration and equivalent plasticity $\gamma$ for discretization with 5k nodes (a,b) and with 50k nodes (c,d).}
\label{fig:neckdeform}
\end{figure}

The load-displacement is plotted and compared to the finite element method, as shown in Fig.\ref{fig:barDispForce}. The results depict that NOM is suitable for large strain J2 plasticity models and can achieve similar results as finite element methods.

\begin{figure}
	\centering
		\includegraphics[width=8cm]{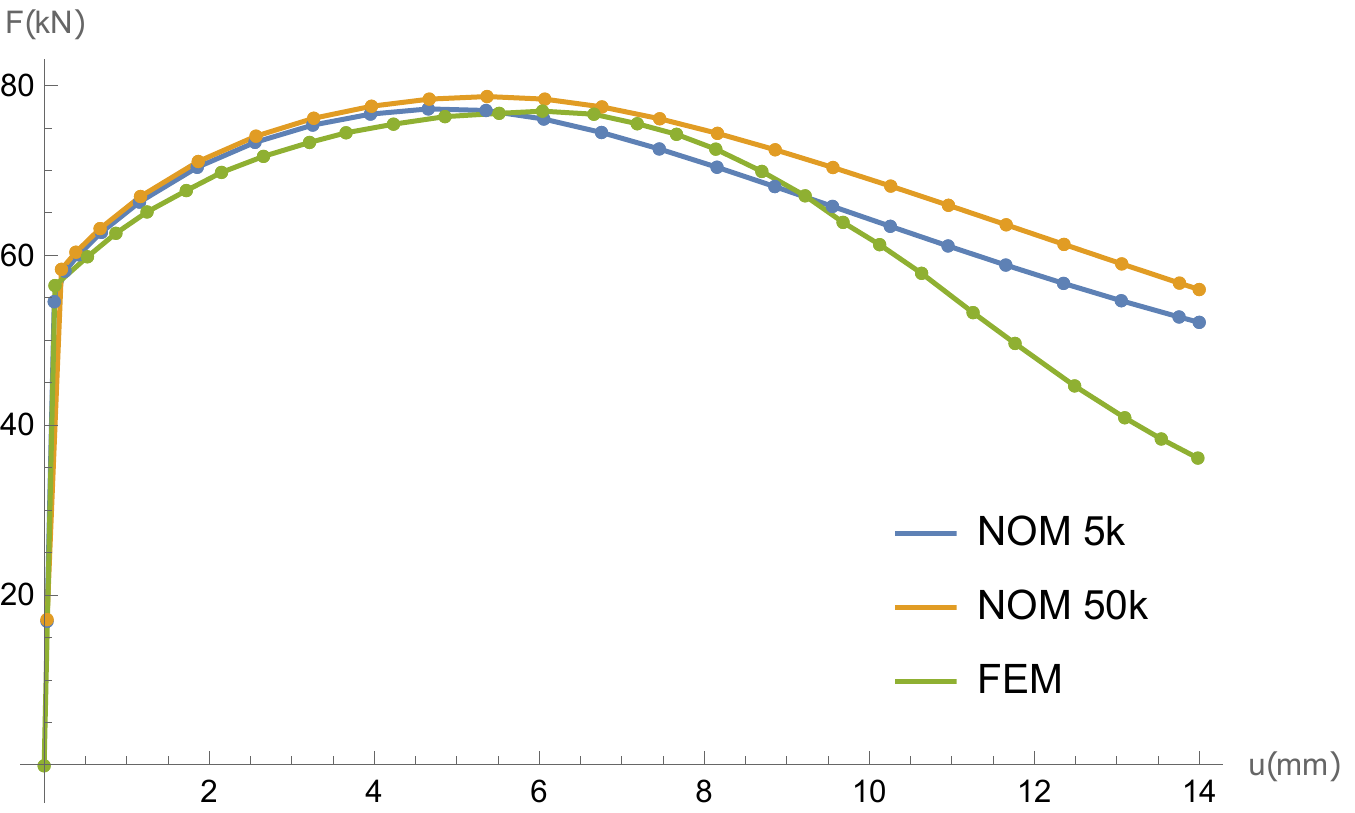}
	\caption{Bar necking: displacement and load curve. The finite element method is referred to as \cite{miehe2002anisotropic}.}
	\label{fig:barDispForce}
\end{figure}

\subsection{Necking of a rod with ductile fracture}
The material parameters, discretization of the 3D rod and the boundary conditions are the same as those in Section \ref{subsec:J2rodnecking}. In order to consider the evolution of ductile fracture, the phase field model is activated. The phase field length scale is selected as the average particle size of $l=1.2$ mm and the critical energy release rate as $G_c=2.7\times 10^{-3}$ kN/mm. We study the resultant reaction forces under the influence of critical release energy rate. For the case of $25 G_c$, the actually deformed particle distribution is given in Fig.\ref{fig:bar5kDeformParticle}. Although the NOM scheme used here is based on nodal discretization, an auxiliary mesh is employed to show the deformation field and other physical quantities such as plasticity and phase field. The deformed configuration, phase field distribution and equivalent plasticity  is shown in Fig.\ref{fig:bar5kDeformPhasePlasticity}. 

\begin{figure}
	\centering
		\includegraphics[width=8cm]{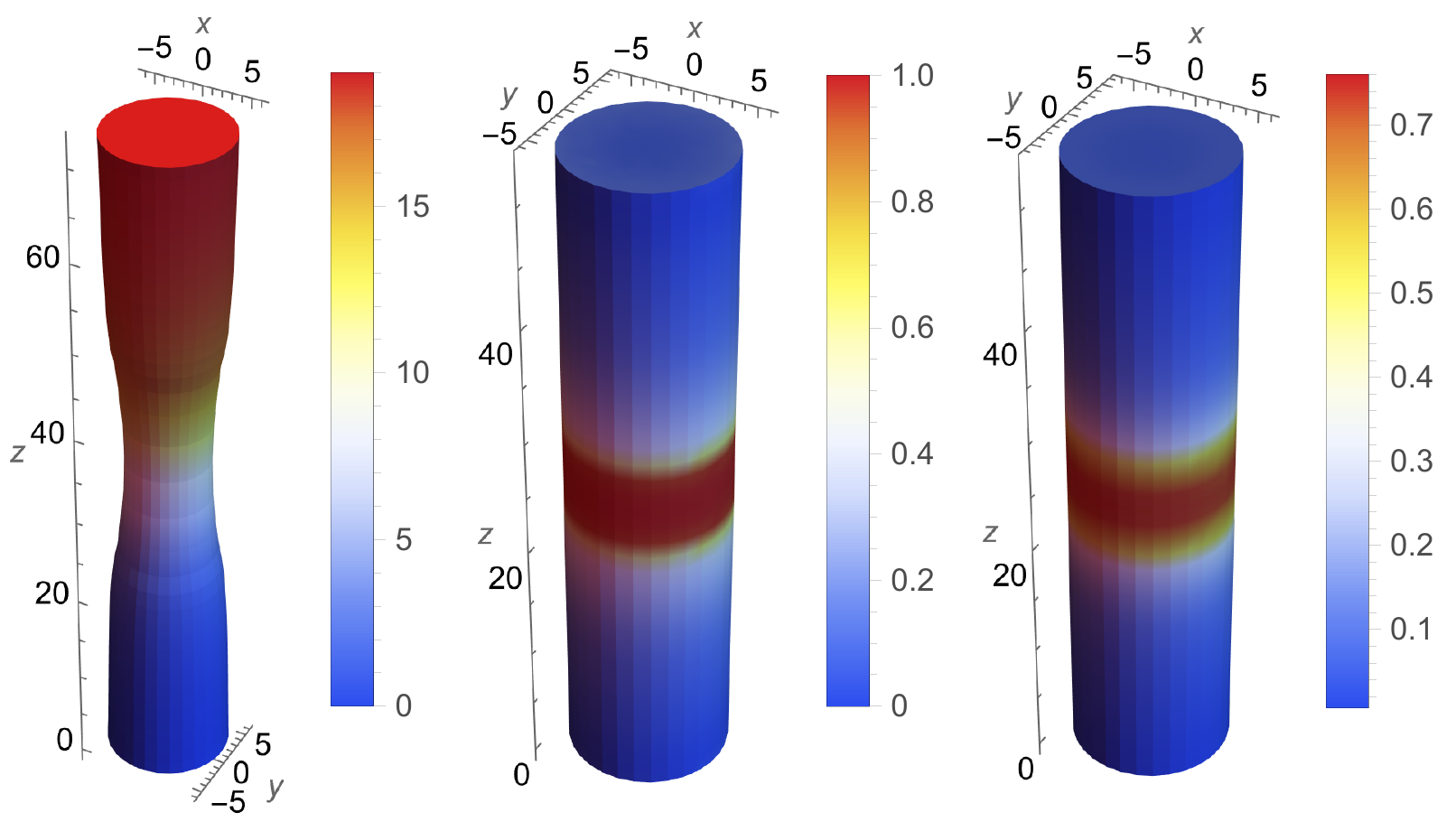}
	\caption{Bar necking with ductile fracture: displacement, phase field and equivalent plasticity.}
	\label{fig:bar5kDeformPhasePlasticity}
\end{figure}

\begin{figure}
	\centering
		\includegraphics[width=8cm]{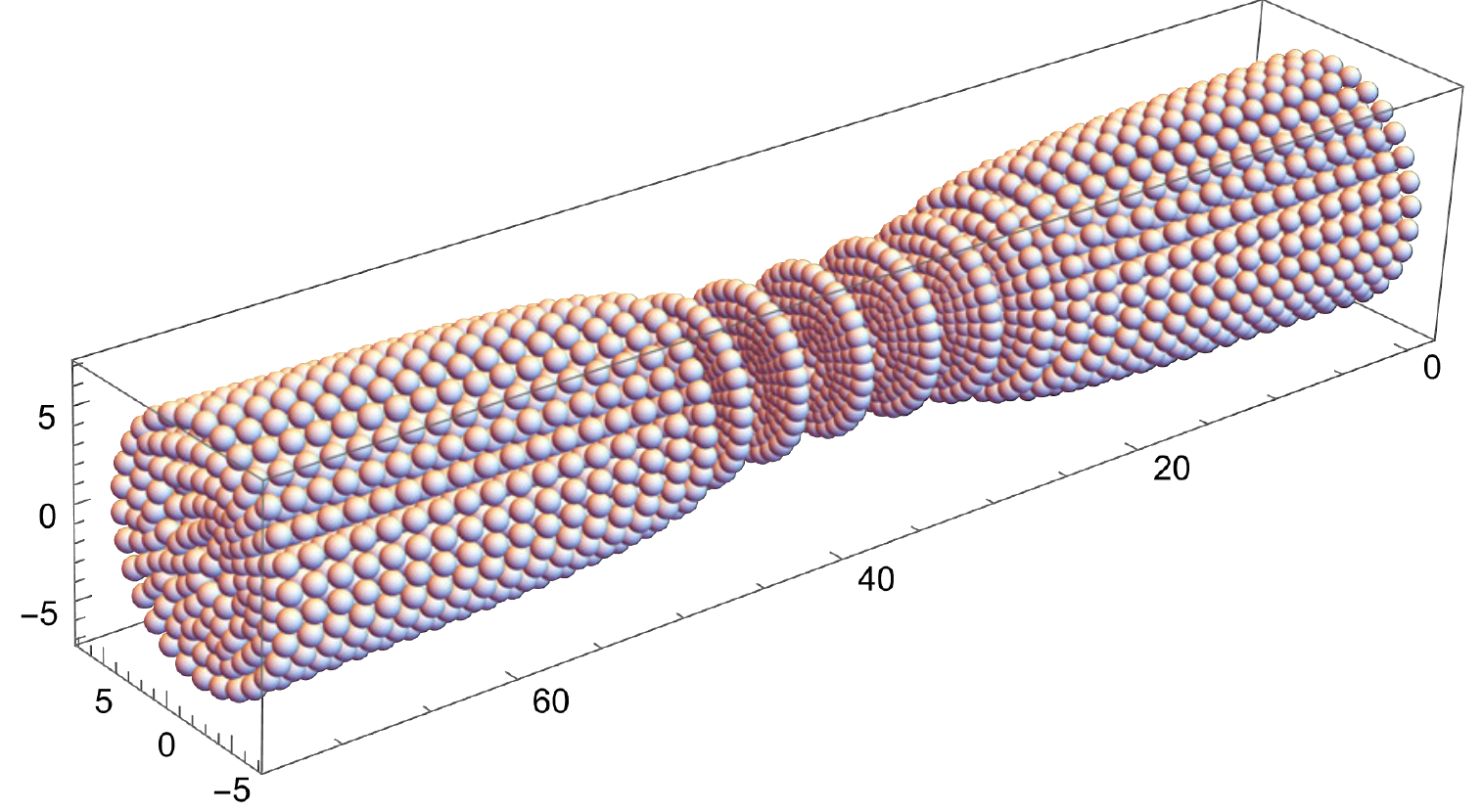}
	\caption{Deformded Bar plotted by particles.}
	\label{fig:bar5kDeformParticle}
\end{figure}

\begin{figure}
	\centering
		\includegraphics[width=8cm]{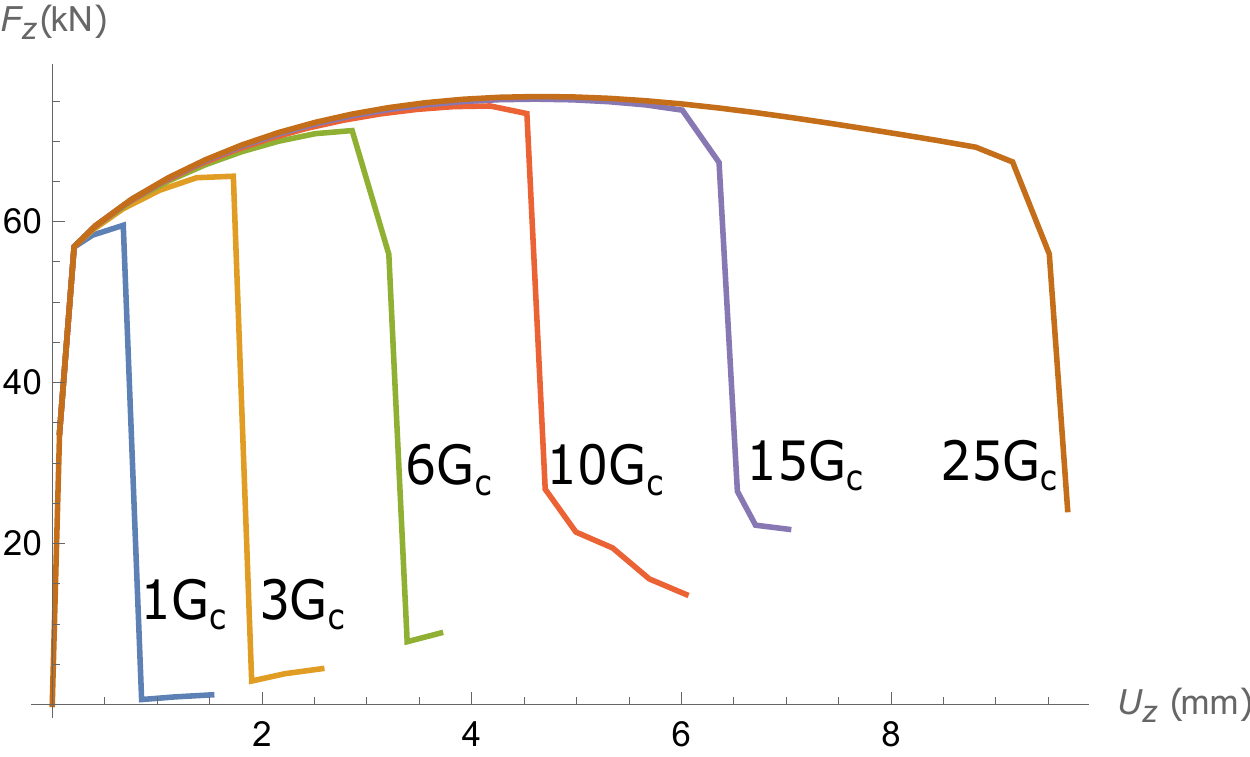}
	\caption{Bar necking with ductile fracture: displacement and load curve.}
	\label{fig:barDispForceDuctile}
\end{figure}

The displacement-load curve for ductile fracture of a rod for different critical energy release rate is depicted in Fig.\ref{fig:barDispForceDuctile}. Conventional J2 plasticity is not easy to describe the damaging process. With the aid of the phase field method, the ductile fracture considering the contribution of phase field deformation can be modeled. It is observed that the ductile fracture differs from the brittle fracture by undergoing firstly a plastic deformation. Since the elastic energy occupies a small part of the total energy, the accumulated plastic energy reaches the threshold value, the damage occurs immediately and the bearing capacity of the rod reduces very fast. With the increase of the critical energy release rate, the critical displacement increases proportionally.

\vspace{-6pt}
\section{Conclusions}
In this paper, we have developed a fintie strain plastic model with phase field method for the modeling of ductile fractures. The model contains several parts: the J2 plasticity model, the consistent tangent stiffness matrix based on matrix functions, the variational derivation of phase field plastic model based on its energy form. The implementation is based on nonlocal operator method, which uses the nonlocal derivative to replace the gradient of shape functions. 

Several numerical examples including the brittle fracture subjected to tensile load and shear load, ductile fracture in 2D and fintie strain J2 plasticity in 3D, and ductile fracture in 3D are presented to show the capability of current scheme.  

\vspace{-2pt}

\vspace{-6pt}
\section*{Acknowledgements}
The first author acknowledges the financial support of the EU project under the title of “Computational Modeling, Topological Optimization and Design of Flexoelectric Nano Energy Harvesters” (ERC COTOFLEXI 802205).

\vspace{-6pt}
\bibliographystyle{plain}
\bibliography{plasticitybib}
\end{document}